\documentclass[journal,draftcls,onecolumn,12pt,twoside]{IEEEtranTCOM}
\usepackage{amsmath}
\usepackage{amsfonts}
\usepackage{url}
\usepackage{array}
\usepackage{bm}
\usepackage{framed}
\usepackage{tikz}
\usepackage{algorithmic}
\usepackage[]{algorithm}
\usepackage{graphicx}
\usepackage{tabularx}
\usepackage{epstopdf}
\usepackage{amsthm}
\usepackage{amssymb}
\usepackage{subcaption}
\usepackage{color} 
\usepackage[numbers,sort&compress]{natbib}
\usepackage[margin=0.99in]{geometry}
\usepackage{multirow}
\usepackage{alltt}
\renewcommand{\baselinestretch}{1.5}
\usepackage[normalem]{ulem}
\usepackage[utf8]{inputenc}
\usepackage{xr}  

\usepackage{pgfplots} 
\pgfplotsset{compat=newest} 
\pgfplotsset{plot coordinates/math parser=false} 
\newlength\figureheight 
\newlength\figurewidth 
\usepackage{bigstrut}
\usepackage{booktabs}
\usepackage{alltt}
\usepackage{multirow,booktabs}
\usepackage{multirow}

\theoremstyle{remark}

\newtheorem{theorem}{Theorem}
\newtheorem{lemma}{Lemma}

\definecolor{mycolor1}{rgb}{0.97255,0.97255,0.97255}%
\newcommand{\distas}[1]{\mathbin{\overset{#1}{\kern\z@\sim}}}%

\def \cU {{\mathcal U}}
\def \D {{\mathbf D}}
\def \s {{\mathbf s}}

\def \d {{\mathbf d}}
\def \m {{\mathbf m}}
\def \X {{\mathbf X}}
\def \Y {{\mathbf Y}}
\def \cF {{\mathcal F}}
\def \cN {{\mathcal N}}
\def \Rn {{\mathbb R}}
\def \lam {{\boldsymbol{\lambda}}}
\def \Lam {{\boldsymbol{\Lambda}}}

\def \u {{\mathbf{u}}}
\def \Z {{\mathbf{Z}}}

\begin{document}
	
	\title{Dynamic Cache Management In \\
		Content Delivery Networks}
	
	\author{Srujan Teja Thomdapu, Palash Katiyar, and Ketan Rajawat}
	\maketitle
	
	\begin{abstract}

The importance of content delivery networks (CDN) continues to rise with the exponential increase in the generation and consumption of electronic media. In order to ensure a high quality of experience, CDNs often deploy cache servers that are capable of storing some of the popular files close to the user. Such edge caching solutions not only increase the content availability, but also result in higher download rates and lower latency at the user. We consider the problem of content placement from an optimization perspective. Different from the classical eviction-based algorithms, the present work formulates the content placement problem from an optimization perspective and puts forth an online algorithm for the same. In contrast to the existing optimization-based solutions, the proposed algorithm is incremental and incurs very low computation cost, while yielding storage allocations that are provably near-optimal. The proposed algorithm can handle time varying content popularity, thereby obviating the need for periodically estimating demand distribution. Using synthetic and real IPTV data, we show that the proposed policies outperform all the state of art caching techniques in terms of various metrics. 
		
	\end{abstract}

\IEEEpeerreviewmaketitle
\section{Introduction} 
As the Internet transitions to the age of on-demand consumption of video and audio content,  developing and maintaining scalable content delivery networks (CDNs) necessitates increasingly sophisticated protocols and algorithms \cite{buyya2008content}. Edge caching is a widely deployed solution that seeks to serve users while ensuring high availability and high quality of experience (QoE) \cite{breslau1999web, barish2000world}. Compared to the central servers, the content stored at the cache servers is small but highly dynamic, consisting primarily of only the ``trending'' files. All file requests are first received by the cache server and the files are served locally if possible. In case of a miss, the request is rerouted to one of the the central repositories, and served at a higher cost in terms of bandwidth usage or latency.   

Caching algorithms ensure high availability of the content close to the users by continuously learning the distribution of content requests. Owing to the limited size of the cache servers, only the most popular files are cached while stale or esoteric content is evicted. Indeed, many caching algorithms are often designated by their respective eviction methods, such as the Least Recently Used (LRU), Least Frequently Used (LFU), First In First Out (FIFO) \cite{maffeis1993cache}, Random \cite{tan2013optimal}, etc. A good caching algorithm not only improves the user experience by ensuring that most of the content is served locally, but also reduces the backhaul traffic, associated bandwidth costs, and the load on the central server. 

Content placement algorithms face complementary challenges pertaining to accurate demand prediction and adaptation to dynamically changing demands. From a theoretical vantage point, the optimal content placement problem can be viewed as a generalization of the facility location problem, that entails solving an integer programming problem \cite{baev2008approximation,drwal2014decomposition}. Formulating the problem within a rigorous utility maximization framework subject to flow and demand constraints has the potential to yield provably optimal storage allocations. However, the resulting integer program cannot be readily solved for realistic CDNs with millions of files. Even for smaller CDNs, the dynamic nature of demands necessitates predicting the demands ahead of time and re-solving the optimization problem every few hours, rendering the approach impractical. 

In practice, eviction-based algorithms are widely used since they are relatively easy to implement and scalable to large CDNs. LRU and its variants, such as meta-LRU and adaptive LRU, are adaptive in nature and can handle time-varying file popularities \cite{li2017accurate}. However, eviction-based algorithms are generally heuristic in nature, cannot be readily generalized to different network settings, and do not offer the flexibility of handling custom cost or utility functions. 
\subsection{Contributions}
This work considers a unified cache content placement problem from the perspective of online optimization. A sequential framework is considered where the demands at each time are revealed after the storage allocation decisions have been made. Different from the integer programming-based content placement approaches, the proposed algorithms are incremental and scalable, and outperform the state-of-the-art eviction methods such as LRU and LFU under both, static and dynamically changing popularities. The proposed framework is flexible and can be applied to different topologies with generic utility and cost functions, while still incurring a computational cost that is linear in the number of files. 

\begin{figure*}
		\centering
		\includegraphics[width=0.8\linewidth,height = 0.4\linewidth]
		{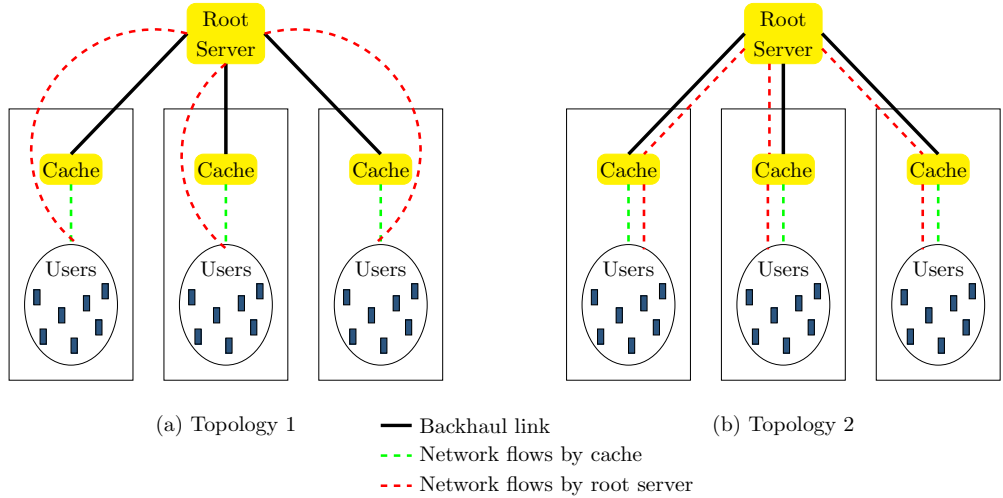}
		\caption{Packet flows} \label{CDN_mainfig}
		\vspace{-1.5cm}
	\end{figure*}

Specifically, a one-shot utility maximization problem is formulated with non-linear cost functions over a finite horizon $T$ and solved in an online manner using the incremental dual descent method. The resulting algorithm is shown to be near-optimal for sufficiently large $T$ while also incurring low update complexity. A key feature of the algorithm is its ability to handle arbitrary and possibly non-stationary demands while still yielding a near-optimal solution. After each update step, the storage allocations are made using the primal iterates so as to ensure feasibility, and different allocation heuristics are proposed. An eviction algorithm is also proposed for the case when the downloaded files are available at the cache for download (see Fig. \ref{CDN_mainfig}).  The performance of the proposed algorithms is extensively tested over synthetic and real IPTV data \cite{elkhatib2014dataset}, and is shown to be superior to state-of-the-art eviction-based algorithms for both static and dynamic popularities. 
\subsection{Related work} \label{rel-work}
Cooperative caching for data management in distributed networks was first studied in \cite{dowdy1982comparative} and has received much attention since then. The performance of a number of caching algorithms has been analyzed through simulations \cite{dahlin1994cooperative}, \cite{fan2000summary}, \cite{ni2005large}, \cite{sarkar2000hint}, and analytical studies \cite{awerbuch1998distributed}, \cite{baev2001approximation}, \cite{baev2008approximation}, \cite{che2001analysis}, \cite{kangasharju2002object}, \cite{korupolu2001placement}, \cite{leff1993replication}. For the most part, these works  focus on latency minimization and  ignore the bandwidth consumption and the associated costs. In contrast, the proposed formulation also takes the backhaul bandwidth consumption into account while designing the algorithm. 

As discussed earlier, content placement algorithms have been formally studied under the aegis of optimization theory. One of the seminal works \cite{baev2001approximation} considered communication cost, storage limits, and demand functions, but show that the resulting problem in generally NP hard. Consequently various approximation techniques have been applied with varying degrees of success \cite{baev2001approximation}, \cite{borst2010distributed}, \cite{valancius2009greening}. Likewise, the caching policies to reduce linear content retrieval costs have been studied in \cite{neglia2018cache} and two new dynamic algorithms are proposed within the static setting. However, these works consider simplistic scenarios with static demands and cannot be generalized to arbitrary network topologies. 

The cache utility maximization problem was first considered in \cite{dehghan2016utility}, where each file is associated with a utility. The idea is inspired from the fact that certain videos with high hit probability have more utility than other videos. The generalized utility driven caching is been proposed and the traditional eviction policies such as LRU and FIFO are modeled as cache utility maximization problems. The utility framework considered here is different and also more general, since the real-world file utilities are time-varying. Recently, the content placement problem for generic networks has been considered from a linear relaxation perspective in \cite{applegate2016optimal}. A mixed integer program is formulated to determine the placement of videos and a decomposition technique is utilized to render the problem solvable for large scale content libraries. In an offline setting, the file placement routine proposed in \cite{applegate2016optimal} is near-optimal and is shown to be orders of magnitude faster than the canonical mixed integer program. However, \cite{applegate2016optimal} cannot be applied to settings where the file popularities are dynamic and unknown in advance. In contrast, the proposed approach builds upon the online gradient descent algorithm allowing low-complexity operation and ability to handle time-varying demands.

The present work considers a classical network topology with a root server and dedicated cache servers. Other topologies have also been considered in the literature. For instance, the network topology in \cite{applegate2016optimal} comprises entirely of inter-connected cache servers and no root server. More generally, \cite{ioannidis2018adaptive} consider an unknown network topology where the source node tries to get the objects from destination node via all possible cache-capable nodes. These and more generic topologies that may arise with a next-generation content centric network \cite{ahlgren2012survey} are not considered here and is left as future work.

Along related lines, the challenge of dealing with non-stationary and time-varying demands has recently been identified in \cite{li2017accurate}. An adaptive LRU algorithm is proposed and shown to outperform classical eviction algorithms in terms of learning efficiency which is a measure of time that algorithm needed to attain the stationary distribution. The optimization-based framework considered here is more general, handling not only time-varying and non-stationary demands, but also different topologies and associated bandwidth limitations. 
	
 The rest of the paper is organized as follows. Sec. \ref{sys-model} details the system model and the problem formulation. The proposed algorithm as well as its performance analysis is provided in Sec. \ref{content_placement}. Next, Sec. \ref{implementation} discusses some of the implementation related aspects of the algorithm. Sec. \ref{simulations} presents detailed simulation results for the proposed algorithms including tests on real data. Finally we conclude the paper in Sec. \ref{conclusion}.

\noindent\textbf{Notation: } Small (capital) bold-faced letters represent column vectors (matrices) while regular font characters represent scalars. Sets are denoted by capital letters in calligraphic font. The cardinality of a set $\mathcal{S}$ is denoted by $|\mathcal{S}|$, and the indicator function $\boldsymbol{1}_{\mathcal{S}}$ is 1 if the condition specified in $\mathcal{S}$ is true, and 0 otherwise. Finally, projection onto the non-negative orthant is denoted by $[\cdot]_{+}$. 
	
	\section{System Model and Problem Formulation }\label{sys-model}
	This section details the system model considered here and formulates the dynamic content placement problem at hand. We begin with a brief overview of the system model. The network comprises of a root server, a number of cache servers $\cN:=\{1, \ldots, N\}$, and a number of users associated with each cache. For simplicity, we consider a setting where each user is associated with a single cache. While only a single root server is considered, the present formulation can be readily extended to the setting with multiple root servers. The file requests from the users are received at the cache servers, and are either accommodated locally or rerouted to the root server, depending on the file availability. Each cache server stores only a subset of the full content library. The goal is to determine the set of files that must be stored at the cache servers. 
	
In order to analyze the flow of content, the network is modeled as a graph whose structure depends on the routes taken by the data packets.  For simplicity, the graph consists of only three kinds of edges or links: backhaul links connected the root and the cache servers, last mile links connecting the cache servers and the users, and optional direct links connecting the root server directly to the users. All three links are present within the first setting, also referred to as the graph topology 1 and depicted in Fig. \ref{CDN_mainfig}. On the other hand, the graph topology 2 consists only of the backhaul and last mile links. Within the second topology, any packets flowing from the root server to the users must pass through the cache servers. The choice of the graph topology is dependent on the caching system available with the CDN. 

Of these, the second topology has been widely used in traditional CDN deployments. Since any file that is downloaded from the central repository passes through the cache, it may be stored at the cache server, if required. The content placement problem therefore reduces to that of deciding whether a particular file is to be kept in the cache or evicted. Online eviction algorithms such as LRU and LFU have been widely studied for this case. In contrast, the first topology has been proposed for next-generation cellular systems that enable edge-caching via mobile small cells. Proactive caching is necessary for such a setting as the caches are populated only at off-peak hours using low-cost backhaul links prior to receiving file requests from the users. During peak hours, most content is required to be downloaded from the cache server itself so that the low-latency and high-cost user-to-root link is activated only sporadically. Different from the existing literature, the present work develops content placement algorithms for both topologies in a unified manner. A generic problem formulation will be presented and nuances pertaining to each topology will be discussed explicitly in Sec. \ref{implementation}. 

\subsection{Network setup}\label{setup}
A generic CDN setup consisting of a root server and $N$ cache servers is considered.  The storage capacity of the $i$-th cache server is denoted by $M_i$ and is measured in bits. The root server contains a library of files indexed as $f \in \cF := \{1, \ldots, F\}$. The size of file $f$ is denoted by $s_f$ and is also measured in bits. The file sizes are collected into the vector $\s \in \Rn_{++}^{F}$.
	
	Time is divided into slots, and time slots are indexed as $t = 1, 2, \ldots$. At time $t$, the set of files stored at the $i$-th cache is denoted by $\cF_i(t) \subset \cF$. Alternatively, the storage vector $\m_i(t)$ for the $i$-th cache is defined as $m_{if}(t) = 1$ for $f\in \cF_i(t)$ and zero otherwise. The storage vector $\m_i(t)$ is viable for the $i$-th cache, if it satisfies $\s^T\m_i(t) \leq M_i$.
	
	Let $\cU_i$ be the set of users that make requests to the $i$-th cache. Since the network capacity and service rates at various servers may be limited, it is assumed that the users in $\cU_i$ can receive at a maximum aggregate rate of $C^c_i$ bits per time slot from the $i$-th cache and $C^r_i$ bits per time slot from the root server. At each time slot $t$, the aggregate demand at the cache $i$ for file $f$ is denoted by $d_{if}(t)$, measured in bits and collected into the vector $\d_i(t) \in \Rn_{+}^{F}$. Given demand $d_{if}(t)$, the service policy is to serve locally if $m_{if}(t) = 1$, and reroute the request otherwise. Assuming that all the demands are subsequently met, the total flow rate out of the $i$-th cache server is given by $\m_i^T(t)\d_i(t)$ and the flow rate from the root server to users in $\cU_i$ is given by $\mathbf{1}^T\d_i(t) - \m^T_i(t)\d_i(t)$, both measured in bits per time slot.
	
	\subsection{Content placement problem}
	Using the notation developed in Sec. \ref{setup}, the content placement problem will now be formulated. Before proceeding however, the traditional approach is first reviewed.
	
\subsubsection{Content placement via integer programming} \label{conint}
When the user demands are known a priori, the optimal content placement involves solving an integer programming problem in $\m_i(t)$; see e.g. \cite{applegate2016optimal}. The idea is to associate appropriate costs with the network flows and find the file placement that minimizes the total cost. For instance, in Fig. \ref{CDN_mainfig}(a) if the cost of using the side link is higher than that of the last mile link, the network cost can be minimized by caching popular content. In the present context, define convex cost functions $\chi_i:\Rn \rightarrow \Rn$ associated with the flows on the last mile links connected to the $i$-th cache server and $\varphi_i:\Rn \rightarrow \Rn$ corresponding to the flows between the root server and the user associated with the $i$-th cache server (see red colored dotted lines in Fig.\ref{CDN_mainfig}). For topology 2 on the other hand, $\varphi_{i}(.)$ represents the cumulative cost function penalizing the flows in the backhaul links and the last mile links connected to the $i$-th cache. Since costs functions can generally be arbitrary, such an approach is significantly more general and offers greater flexibility as compared to the traditional eviction-based policies. As an example, the squared penalty $\chi_i(x) = c_1x^2+c_2$ engenders a base cost in addition to a squared network usage cost that discourages very high usage at any single server, thereby avoiding the possibility of buffer overflows or dropped requests. Another example is that of simple linear costs augmented with a strongly convex regularizer, e.g., $\chi_i(x) = wx + \tfrac{1}{2}x^2$, that imposes a per-bit charge of $w$ units, in addition to penalizing large file downloads. Finally, the Kleinrock's average delay function for a bottleneck link with capacity $C_i$ is given by $\chi_i(x) = x/(C_i-x)$ for $0< x < C_i$.  In contrast, delays cannot generally be directly controlled by tuning the parameters of eviction-based policies. 

The content placement problem at each time $t$ can therefore be formulated as the following integer program:
	\begin{subequations}\label{cdnint}
	\begin{align}
	\min_{\{\m_i(t) \in \{0,1\}^F\}_{i}} &\sum_{i,f} \chi_i\Big(m_{if}(t)d_{if}(t)\Big) +\sum_{i,f}\varphi_i\Big(d_{if}(t)(1 - m_{if}(t))\Big) \label{obji}\\
	\text{s. t. } & \m_i^T(t)\d_i(t) \leq C^c_i, \hspace{0.5cm} \mathbf{1}^T\d_i(t) - \m^T_i(t)\d_i(t) \leq C^r_i \hspace{0.25cm} \forall i\in\mathcal{N}\label{crc}\\
	& \s^T\m_i(t) \leq M_i \hspace{6.3cm} \forall i\in\mathcal{N}\label{size}
	\end{align}
	\end{subequations}
That is, the goal is to find the placement vectors $\{\m_i(t)\}$ that result in the minimum cost at each time $t$. The constraints \eqref{crc} ensure that the cumulative flows do not exceed the respective link capacities while \eqref{size} ensures that at most $M_i$ bits are stored at the $i$-th cache. 

Observe that \eqref{cdnint} is an integer program and is generally difficult to solve at every time instant, especially when the content library consists of a large number of files. More importantly however, in real-world settings, the user demands $\d_t(t)$ are not known prior to the placement. The first topology in particular requires the placement to be carried out periodically and possibly several hours before the user demands are revealed. Even in topology 2, only the individual file demands corresponding to a few users (instead of the entire demand vector) are known before the placement decision has to be made. Such a setting is instead reminiscent of the online learning framework \cite{ribeiro2010ergodic}, \cite{wang2011resource} and therefore cannot be tackled using classical offline optimization algorithms. 
	
\subsubsection{Relaxed Problem} \label{relax}
In order to address the issues with the integer programming formulation, two relaxations are introduced. For each $i$ and $t$, let $x_{if}(t) = m_{if}(t)d_{if}(t)$ and $y_{if}(t) = (1-m_{if})d_{if}(t)$ be the variables denoting the anticipated flows to the users in $\cU_i$ by the cache and the root server respectively. The two sets of variables are collected into $N \times F$ real matrices $\X(t)$ and $\Y(t)$ respectively. The integer constraint on $m_{if}(t)$ is relaxed by allowing $x_{if}(t)$ and $y_{if}(t)$ to take non-negative real-values. Further, the relationship between the anticipated flows is relaxed to hold in the average sense only. In other words, the anticipated flows meet the demand on an average over a horizon of $T$ time slots, i.e.,
	\begin{align}\label{relavgdmd}
	\frac{1}{T}\sum_{t=1}^T \left(x_{if}(t) + y_{if}(t) - d_{if}(t)\right) &= 0.
	\end{align}
Such a relaxation is necessary since the original relationship $x_{if}(t) + y_{if}(t) = d_{if}(t)$ cannot be imposed for practical systems. Instead, the flows will be estimated ahead of time in anticipation of the future demands. The anticipated flows will subsequently be used to make decisions regarding content placement. Note that the actual flows at the time of content delivery will likely be different and would necessarily adhere to the exact user demands. However, the introduction of the anticipated flows serves as a mechanism that allows us to keep track of the placement and content popularity. Such 'over-the-horizon' constraints are commonly used in the context of online resource allocation settings where the network states are revealed in a sequential manner; see e.g. \cite{wang2011resource}, \cite{ribeiro2010ergodic}. 
	
 These two relaxations allow us to eliminate the storage variable $\m_i(t)$, and pose the problem in terms of the anticipated flows $x_{if}(t)$ and $y_{if}(t)$, over a horizon of $T$ time slots,
	\begin{subequations}\label{cdnlin}
		\begin{align}
		\min_{\X(t),\Y(t) \in \Rn_{+}^{N\times F}}& \frac{1}{T}\sum_{t=1}^T \sum_{i,f}  \bigg( \chi_i\Big( x_{if}(t)\Big)  +  \varphi_i\Big( y_{if}(t)\Big) \bigg) \label{cdnlin1}\\
		\text{s. t. } & \sum_{f} x_{if}(t) \leq C^c_i, \hspace{1cm}  \sum_{f} y_{if}(t) \leq C^r_i \hspace{2cm} \forall i\in\mathcal{N} \label{xycap}\\
		& \frac{1}{T}\sum_{t=1}^T \Big(x_{if}(t)+y_{if}(t)-d_{if}(t)\Big) = 0 \hspace{1cm} \forall i\in\mathcal{N}, \forall f\in\mathcal{F} \label{constraint1}
		\end{align}
	\end{subequations}

Clearly, {problem \eqref{cdnlin}} can be solved in an offline fashion if the demands are known ahead of time. In the present case however, we put forth an online algorithm that allows us to take actions in a sequential manner and will be shown to be asymptotically near-optimal. The overall idea of the algorithm is summarized as follows
\begin{itemize}
	\item determine the anticipated flows $\X(t)$ and $\Y(t)$ at the start of the $t$-th time slot and before the current demands $\D(t)$ are revealed;
	\item make valid placement decisions $\m_i(t)$ based on $\{\X(t),\Y(t)\}$ while adhering to storage constraints \eqref{size}; and
	\item serve the user demands as usual, i.e., locally if the file is available at the cache and via the root server otherwise. 
\end{itemize}
The first step that yields the anticipated flows ahead of time is the most important and will be responsible for correctly tracking the popularity of each file and adhering to the network capacity constraints \eqref{crc}. The design of an online algorithm will be discussed in the subsequent section. The second step is necessary since the anticipated flows were not designed to adhere to the storage constraints, that must be imposed explicitly; see Sec. \ref{proposed_techniques}. In the third step, the decisions of the CDN depend on the received demand vector. If the network is sufficiently over-provisioned, the capacity constraints \eqref{crc} are always satisfied from the first step. 

\section{Content Placement via Dual Ascent} \label{content_placement}
This section develops the incremental dual ascent algorithm (Sec. \ref{incremental}) and provides theoretical guarantees on its performance (Sec. \ref{asym_prop}).  The implementation details including the proposed and existing content placement algorithms are provided in Sec. \ref{implementation}.

\subsection{Incremental Dual Ascent}\label{incremental}
Since the problem at hand is high-dimensional, first order methods are well motivated for an efficient and scalable implementation.	Associate dual variables $\{\lambda_{if}\}$ with the constraints in \eqref{constraint1} and denote the collected dual variable matrix as $\Lam \in \Rn^{N \times F}$. The Lagrangian can therefore be written as
	\begin{align*}
	L(\{\X(t), \Y(t)\}_t, \Lam) :=&\frac{1}{T}\sum_{t=1}^T \sum_{i,f}\bigg[\chi_i\Big(x_{if}(t)\Big) - \lambda_{if}x_{if}(t) +{\varphi_i\Big(y_{if}(t)\Big)}-\lambda_{if}y_{if}(t)+\lambda_{if}d_{if}\bigg]	
	\end{align*}
	The separable nature of the Lagrangian allows us to write the dual function as $\varrho(\Lam) = \frac{1}{T}\sum_{t=1}^T\varrho_t(\Lam)$, where
	\begin{align*}
	\allowdisplaybreaks
	\varrho_t(\Lam) = \min_{\X(t),\Y(t) \in \Rn_{+}^{N \times F}}&  \sum_{i,f}\bigg[\chi_i\Big(x_{if}(t)\Big)-\lambda_{if}x_{if}(t) + \varphi_i\Big(y_{if}(t)\Big)-\lambda_{if}y_{if}(t) +\lambda_{if}d_{if}(t)\bigg] \\
	\text{s. t. }  &\sum_{f} x_{if}(t) \leq C^c_i,  \hspace{1cm}\sum_{f} y_{if}(t) \leq C^r_i \hspace{2cm} \forall i\in\mathcal{N}
	\end{align*}
	Finally, the dual problem becomes that of maximizing $\varrho(\Lam)$ with respect to $\Lam$. Since the dual function is expressible as a sum over $t$, it is possible to use the incremental gradient method to solve the dual problem \cite{blatt2007convergent}. The resulting updates take the following form for each cache $i = 1, 2, \ldots, N$,
	\begin{subequations}\label{ida}
		\begin{align}
		\{\hat{x}_{if}(t)\}_f &= \arg\min_{\{x_{f} \geq 0\}}  \sum_{f=1}^F \chi_i(x_f) - \lambda_{if}(t)x_f \hspace{2cm}\text{s. t. } \sum_{f=1}^F x_f \leq C^c_i \label{xup}\\
		\{\hat{y}_{if}(t)\}_f &= \arg\min_{\{y_{f} \geq 0\}}  \sum_{f=1}^F \varphi_i(y_f) - \lambda_{if}(t)y_f  \hspace{2.1cm}\text{s. t. }\sum_{f=1}^F y_f \leq C^r_i \label{yup}\\
		\lam_{if}(t+1) &= \lam_{if}(t) - \mu \Big(\hat{x}_{if}(t)+\hat{y}_{if}(t)-d_{if}(t)\Big) \hspace{2cm}\forall f\in\mathcal{F} \label{lup}
		\end{align}
	\end{subequations}
	where $\mu>0$ is the step size and the algorithm is initialized with an arbitrary $\lam_{if}(0)$. Here, the updates in \eqref{xup}-\eqref{yup} constitute the primal updates while \eqref{lup} is the dual update. Observe that the primal update subproblems are convex and can generally be solved via interior point methods at complexity $\mathcal{O}(F^3)$ for each $i \in \mathcal{U}$. However, in practical CDNs with hundreds of thousands of files, even cubic complexity may be prohibitive. To this end, we propose a simpler update algorithm that can solve \eqref{xup}-\eqref{yup} with near linear complexity. The proposed algorithm requires the following assumption.
	\begin{enumerate}
		\item[\textbf{A0.}] The penalty functions $\chi_i(u)$ and $\varphi_i(v)$ are strictly  convex and differentiable over the ranges $u \in [0,C^c_i]$ and $v \in [0,C^r_i]$ respectively.
	\end{enumerate}
	It is remarked that such a requirement is not very restrictive and is satisfied by most cost functions such as the square penalty function introduced earlier. Moreover, most cost functions can generally be approximated with ones that satisfy (\textbf{A0}). 
	
	In order to simplify the exposition, consider the following general problem that subsumes \eqref{xup}-\eqref{yup}:
	\begin{align}\label{pg}
	\hat{\u} &= \arg\min_{\u\geq 0}\sum_{f = 1}^F h(u_f)  - \lam^T\u & \text{s. t. }& \mathbf{1}^T\u \leq C 
	\end{align}
	where $[\lam]_f = \lambda_{f}$, $[\u]_f = u_f$, and $h(\cdot)$ is a strictly convex and differentiable function. Clearly, both \eqref{xup} and \eqref{yup} are of the same form as \eqref{pg}. Since $h(\cdot)$ is strictly convex, $h'(\cdot)$ is monotonically increasing and $h'^{-1}(\cdot)$ is a well-defined function. The following lemma provides the necessary results for solving \eqref{pg} at low complexity. The proof of Lemma \ref{primsol} is provided in Appendix \ref{aprimsol}.
	
	\begin{lemma}\label{primsol}
		Defining $g_f(z) = h'^{-1}(\lambda_f-z)$, the following results hold under (\textbf{A0}):
		\begin{enumerate}
			\item[(a)] If $g_f(0) \leq 0$ then it holds that $\hat{u}_f = 0$.
			\item[(b)] If $\sum_f [g_f(0)]_{+} > C$, then it holds that $\sum_f \hat{u}_f = C$.
			\item[(c)] The solution to \eqref{pg} is given by 
			\begin{align}\label{pgsol}
			\hat{\u}_f = \begin{cases} [g_f(0)]_{+} & \text{if }\sum_f [g_f(0)]_{+} \leq  C \\
			[g_f(\hat{\upsilon})]_{+} &\text{otherwise} \end{cases}
			\end{align}
			where $\hat{\upsilon} \in [0,\max_f\lambda_f - h'(0)]$ is the solution to $\sum_f [g_f(\hat{\upsilon})]_{+} = C$. 
		\end{enumerate}
	\end{lemma}
	
	While the solution in \eqref{pgsol} is not in closed form, it can be calculated using the bisection routine over $\hat{\upsilon} \in [0, \max_f\lambda_f - h'(0)]$. The bisection algorithm incurs a complexity of $\mathcal{O}(F\log(1/\delta))$ to yield a solution that is $o(\delta)$ accurate. In other words, given the desired accuracy (say $\delta = 10^{-6}$), the proposed algorithm for solving \eqref{xup}-\eqref{yup} incurs a complexity that only grows linearly with the number of files $F$. It is remarked that the primal updates are only required to be carried out once per update interval, that typically ranges from few hours to a day; see Sec. \ref{simulations}. The full algorithm used for solving \eqref{pg} is summarized in Algorithm \ref{opt_flow}.
	
	\begin{algorithm}
		\caption {Algorithm to solve \eqref{pg}}
		\begin{algorithmic}[1]
			\STATE \textbf{Arrange} $\{\lambda_f\}_f$ in ascending order such as $\lambda_m \leq \lambda_n$, for any $m<n$. 
			\STATE \textbf{Initialize} $\{\hat{u}_f\}_f=\{\left[g_f(0)\right]_+\}_f$, where $g_f(z) = h'^{-1}(\lambda_f-z)$
			\STATE \textbf{While} $\sum_f\hat{u}_f> C$
			\STATE\hspace{3mm} $k=|\{f:\hat{u}_f=0\}|$
			\STATE\hspace{3mm} \textbf{Define} $\rho(.)=\sum_fg_f(.)\mathbf{1}_{\{\hat{u}_f>0\}}$
			\STATE\hspace{3mm} \textbf{if} $\rho\left(\lambda_{k+1}-h'(0)\right)\leq C$
			\STATE\hspace{6mm} \textbf{for} $f \in \{f:\hat{u}_f>0\}$
			\STATE\hspace{9mm} $\hat{u}_f=g_f\left(\rho^{-1}(C)\right)$.
			\STATE\hspace{6mm} \textbf{end}
			\STATE\hspace{3mm} \textbf{else}
			\STATE\hspace{6mm} $\hat{u}_{k+1}=0$.
			\STATE\hspace{3mm} \textbf{end}
			\STATE \textbf{end}
		\end{algorithmic}
		\label{opt_flow}
	\end{algorithm}
	
	It is remarked that an online algorithm such as \eqref{ida} cannot yield the optimal solution to \eqref{cdnlin} over a finite horizon. Instead, the subsequent section establishes that for strongly convex costs, the proposed incremental algorithm yields an $\epsilon$-optimal solution to \eqref{cdnlin} after $\mathcal{O}(1/\epsilon^2)$ updates. 

	\subsection{Near-optimality of the incremental update algorithm }\label{asym_prop}
	For ease of analysis and in order to obtain meaningful results, this section introduces certain regularity assumptions on the problem parameters. Although the assumptions required here are stronger than (\textbf{A0}), they are also fairly standard and are simply required to eliminate some of the pathological choices of demands and costs.
	\begin{enumerate}
		\item[\textbf{A1.}] The network capacity is sufficiently large so as to satisfy the aggregate demands, i.e., $\sum_f d_{if}(t) \leq C_i^c+C_i^r$.
		\item[\textbf{A2.}] The penalty functions $\chi_i(u)$ and $\varphi_i(v)$ are strongly convex and Lipschitz over the region, i.e., for $0 \leq x_2 \leq x_1 \leq C^c_i$ and $0 \leq y_2 \leq y_1 \leq C^r_i$, it holds that
		\begin{subequations}\label{a3}
			\begin{align}
			&m_\chi ( x_1 - x_2 )\leq \chi_i'(x_1) - \chi_i'(x_2) \leq L_\chi(x_1 - x_2)  \\
			&m_\varphi ( y_1 - y_2 )\leq \varphi_i'(y_1) - \varphi_i'(y_2)\leq L_\varphi(y_1 - y_2)
			\end{align}
		\end{subequations}
		where the parameters $m_\chi$, $m_\varphi$, $L_\chi$, and $L_\varphi$ are all positive. 
	\end{enumerate}
	
	Of these, Assumption (\textbf{A1}) holds if the network is over-provisioned. Note that the problem formulation in \eqref{cdnint} is not applicable to the scenario when the network capacity is below the peak demand and some of the users are denied service. On the other hand, Assumption (\textbf{A2}) can be satisfied via appropriate choice of the cost functions. The strong convexity and Lipschitz requirement is standard in the context of convex optimization algorithms (see e.g.  \cite{chen2017stochastic}), and imply quadratic upper and lower bounds on the penalty functions. It is remarked that the squared penalty function introduced earlier satisfies (\textbf{A2}). Likewise, it is possible to use a linear cost function augmented with a regularization term. Similarly, the cost function $\chi_i(x) = kx^2/(C_i-x) + k_i$, where the first term penalizes the average number of packets in an M/M/1 queue with service rate $C_i$ \cite{kleinrock1975queueing}, also satisfies (\textbf{A2}) over the interval $[0,x_{\max}]$ with $x_{\max}<C_i$.
	
	The asymptotic results in this section also require that the initial dual iterates $\{\lambda_{if}(0)\}$ are selected such that the following two conditions hold:
	\begin{subequations}\label{lambdalimits}
		\begin{align}
		\lambda_{if}(0) &\geq \min\{\chi'_i(0),\varphi_i'(0)\} & \forall ~ i,f\label{minlambda}\\
		\sum_{f=1}^F \lambda_{if}(0) &\leq \Delta &\forall~i \label{maxlambda}
		\end{align}
	\end{subequations}
	where $\Delta:= \max\{L_{\chi},L_{\varphi}\}(C^c_i+C^r_i) + F\max\{\chi'_i(0),\varphi_i'(0)\}$. For the subsequent discussion, let $L:=\max\{L_{\chi},L_{\varphi}\}$, and $m:=\min\{m_{\chi},m_{\varphi}\}$. Before stating the main result of this section, the following intermediate lemma regarding the boundedness of $\{\lambda_{if}(t)\}$ is first established. 
	
	\begin{lemma}\label{lambda_bound}
		Under (\textbf{A1})-\textbf{(A2)} and for $0< \mu < m$, the updates in \eqref{ida}, when initialized in accordance with \eqref{lambdalimits}, adhere to the following limits for all $t\geq 1$,
		\begin{subequations}\label{boundlambdat}
			\begin{align*}
			\lambda_{if}(t) \geq \min\{\chi'_i(0),\varphi_i'(0)\} \hspace{0.5cm}\forall ~ i \in \mathcal{N}, f \in \mathcal{F}, \hspace{2cm}\sum_{f=1}^F \lambda_{if}(t) \leq \Delta \hspace{0.5cm}\forall ~ i \in \mathcal{N} 
			\end{align*}
		\end{subequations}
	\end{lemma}
	Lemma \ref{lambda_bound} ensures that the dual iterates, if initialized appropriately and for $\mu$ small enough, continue to stay within certain limits for all $t\geq 1$. Interestingly, the result holds regardless of the demands $\{d_{if}(t)\}$, as long as the bound in (\textbf{A1}) is satisfied. The proof of Lemma \ref{lambda_bound} is provided in Appendix \ref{alamb}. 
	
	For the final result, associate Lagrange multipliers $\alpha_i(t)$ and $\beta_i(t)$ with the constraints in \eqref{xycap} respectively. Likewise, associate constraints $\xi_{if}(t)$ and $\zeta_{if}(t)$ with non-negativity constraints $x_{if}(t)\geq 0$ and $y_{if}(t) \geq 0$ in \eqref{cdnlin}, respectively. Finally, associating Lagrange multiplier $\nu_{if}$ with \eqref{constraint1}, the full Lagrangian can be written as
	\begin{align}
	L_{T}&(\{\X(t), \Y(t), \bm{\alpha}(t), \bm{\beta}(t), \bm{\xi}(t), \bm{\zeta}(t)\}_{t=1}^T, \bm{\nu}) = \frac{1}{T}\sum_{t=1}^T \sum_{i,f} \chi_i(x_{if}(t)) + (\alpha_i(t)-\xi_{if}(t)-\nu_{if})x_{if}(t) \nonumber\\
	&\hspace{0.45cm}+ \varphi_i(y_{if}(t)) + (\beta_i(t)-\zeta_{if}(t)-\nu_{if})y_{if}(t)- C_i^c\alpha_i(t) - C_i^r\beta_i(t) + \nu_{if}d_{if}(t) 
	\end{align}
	where the bold quantities denoted the corresponding collected variables for all $i$ and $f$. Let $\Z(t)$ collect the variables $\{\X(t), \Y(t), \bm{\alpha}(t), \bm{\beta}(t), \bm{\xi}(t), \bm{\zeta}(t)\}_{t=1}^T$.
	
	\begin{theorem} \label{thm}
		Under (\textbf{A1})-\textbf{(A2)}, initialization in accordance with \eqref{lambdalimits}, and for any $0< \mu < m$, the updates in \eqref{ida} yield flows that are near-optimal for \eqref{cdnlin} in the following sense:
		\begin{align}
		\left|\frac{1}{T}\sum_{t=0}^{T-1}\left(\hat{x}_{if}(t) + \hat{y}_{if}(t) - d_{if}(t) \right)\right|  & \leq \mathcal{O}\left(\frac{1}{\mu T}\right) \label{thmpf}\\
		\left\|\nabla_{\{\X(t),\Y(t)\}_{t=1}^T}  L_T(\{\hat{\Z}(t)\}_{t=1}^T, \frac{1}{T} 
		\sum_{t=1}^T \lambda_{if}(t))\right\|_2  & \leq \mathcal{O}\left(\frac{1}{\sqrt{T}}\right) \label{thmlt}
		\end{align}
	\end{theorem}
		
	    \begin{algorithm}
		\caption {Content Placement Algorithm}
		\begin{algorithmic}[1]
			\STATE {\bf Initialize} $\boldsymbol{\lambda}=\boldsymbol{\lambda}(0)$ as in \eqref{lambdalimits}, $t=0$, $\D^{N\times F}=\textbf{0}$.
			\STATE \textbf{Input} Cache update interval $T_{\text{up}}^v$ and the dual update interval $T^{\lam}_{\text{up}}$
			\STATE \textbf{Repeat $t \geq 0$}
			\STATE\hspace{2mm} \textbf{if} $t$ $\%$ $T^v_{\text{up}}=0$
			\STATE\hspace{4mm} $\textbf{M}(t) \leftarrow cache(\X(t),\Y(t))$ \STATE\hspace{4mm} Update cache storage $\{\mathcal{F}\}_i$.
			\STATE\hspace{2mm} \textbf{end}
			\STATE\hspace{2mm} $t$ $\leftarrow$ $t+1$
			\STATE\hspace{2mm} Accumulate the demand matrix $\D=\D+\D(t)$.
			\STATE\hspace{2mm} \textbf{if} $t$ $\%$ $T^{\lam}_{\text{up}}=0$
			\STATE\hspace{4mm} Find the anticipated flows as in \eqref{xup}, \eqref{yup}.
			\STATE\hspace{4mm} $\boldsymbol{\lam}=\boldsymbol{\lam}-\mu\left(\X(t-1)+\Y(t-1)-\D\right)$ as in \eqref{lup}.
			\STATE\hspace{4mm} $\D=0$.
			\STATE\hspace{2mm} \textbf{end}
		\end{algorithmic}
		\label{content_placement_IPTV}
	\end{algorithm}	
	The proof of Theorem \ref{thm} is deferred to Appendix \ref{athm}. Theorem \ref{thm} establishes the near-optimality of the anticipated flows obtained from the updates in \eqref{ida} with respect to the relaxed problem in \eqref{cdnlin}. In particular, it follows from Theorem \ref{thm} that the anticipated flows satisfy the KKT conditions of \eqref{cdnlin} approximately with accuracy bounded by $1/\sqrt{T}$. In other words, the incremental algorithm requires at least $1/\epsilon^2$ updates in order to yield an $\mathcal{O}(\epsilon)$ accurate solution to \eqref{cdnlin}. The result is interesting and differs from the related results on stochastic optimization, where stationarity assumptions are often required for the process $\{d_{if}(t)\}$. Indeed, relaxing the stationarity assumption in the Markov decision theoretic, backpressure, or stochastic approximation frameworks is not straightforward, and has only been attempted for special cases. On the other hand, the current result hold for any arbitrary sequence of user demands. 
	
	The flexibility afforded with respect to the temporal variations in the demands is also important in practice. For instance, the demands $\{d_{if}(t)\}_{t\geq 1}$ are usually not i.i.d. but exhibit temporal decay and correlation. In particular, the content popularity often decreases over time \cite{elkhatib2014dataset}. Similarly. it was observed in \cite{cha2009analyzing} for instance, that the demands often exhibited a diurnal pattern, reaching peak values at specific times of the day.
	
	\subsection{Implementation and intuition} \label{intu}
The complete implementation of the content placement algorithm is summarized in Algorithm \ref{content_placement_IPTV}. Demands are accumulated over a duration $T_\text{up}^{\lam}$ and are used to populate the matrix $\D(t)$. At every update interval, the anticipated flow variables $\{\hat{x}_{if}, \hat{y}_{if}\}$ are obtained from Algorithm \ref{opt_flow} using the current value of $\lambda_{if}$, which is updated next. The anticipated flows are also used to update the storage allocation vectors $\{\m_i(t)\}$ as specified in line 5 of Algorithm \ref{content_placement_IPTV}. The function $cache(\X(t),\Y(t))$ denotes the use of one of the placement policies detailed in Sec. \ref{proposed_techniques}, namely, Top-$X$, Least-$X$, Least-$X_{\text{th}}$, and Least-$X_f$. Cache update interval is denoted as $T_{\text{up}}^v$ (see line 4 Algorithm. \ref{content_placement_IPTV}). Intuitively, the placement policies are rounding algorithms that convert the solution of the relaxed problem in \eqref{cdnlin} to suboptimal but feasible solutions to the original placement allocation problem in \eqref{cdnint}. The efficacy of the placement policies will be tested in Sec. \ref{simulations}, where it will be established that placements made in accordance with the solutions obtained from (4) are significantly better than those made according to various heuristics  used in the literature.

Towards obtaining further intuition regarding the working of the algorithm, it may be useful to think of $\lam$ as shadow prices associated with not meeting the demand. Indeed, as evident from \eqref{lup}, the prices increase when the demand is not met by the anticipated flows at any time $t$ or equivalently when $x_{if}(t)+y_{if}(t)-d_{if}(t)$ is negative. In the primal update steps \eqref{xup}, \eqref{yup}, the anticipated flows are adjusted depending on the shadow prices are high or low. For instance, if the demands are stationary, the anticipated flows continue to follow them such that the difference $x_{if}(t)+y_{if}(t)-d_{if}(t)$ does not become very large.

As a simple example, consider a file $f$ whose demand from a user $i$ is zero for all $t$. For such a file, if $\lambda_{if}(0) = 0$, it will continue to be zero for all time $t$. Indeed, even if the initialization is different, the dual variables will approach zero, since $\hat{x}_{if}(t)$ and $\hat{y}_{if}(t)$ will be strictly positive whenever $\lambda_{if}(t) > 0$. Conversely, it is possible to argue that if a particular $d_{if}(t)$ remains high for all $t$, so will $\lambda_{if}(t)$. Extending the argument, the dual variable $\lambda_{if}(t)$ ``follows'' the demands $d_{if}(t)$, albeit its tracking ability is limited via the step size $\mu$. Consequently, the anticipated flows $\{\hat{x}_{if}(t), \hat{y}_{if}(t)\}$, which are monotonic functions of $\lambda_{if}(t)$, also track the popularity of files. Moreover, intermittent spikes or dips in the demands have minimal effect on the evolution of $\lambda_{if}(t)$ owing to the averaging effect inherent to the dual algorithm. 

The interpretation provided here will form the basis of the proposed content placement algorithm. The observation that the anticipated flows represent the averaged popularity of the files suggests they can be used to perform content placement in an online fashion. The next section builds upon the existing caching techniques to develop improved placement algorithms.   
	
\section{Storage allocation} \label{implementation}
This section cements the link between original problem in \eqref{cdnint}, and the relaxed problem in \eqref{cdnlin}. We use the anticipated flows obtained from \eqref{ida} to determine the various storage allocation policies. Within this context, recall that the anticipated flows may not necessarily meet the demands at every time instant, but only on an average. On the other hand, the storage allocations obtained here will be feasible, but not necessarily optimal for \eqref{cdnint}. We begin with briefly reviewing some of the existing eviction techniques.
	
	\subsection{Existing eviction techniques} \label{Existing_cache_techniques}
Eviction-based content placement algorithms apply to topology 2 and are generally oblivious to the bandwidth usage. Upon receiving a request, the cache server checks if the file is available locally or not. In case of a cache miss, the file is downloaded from the central repository, served to the user, and stored at the cache server. When the cache is full, various eviction algorithms are used, as detailed next.

The \emph{least recently used} (LRU) and the \emph{least frequently used} (LFU) eviction techniques are the two commonly used techniques. As their names suggests, if the cache is full, the least recently used (respectively, least frequently used) file is evicted. The \emph{random replacement} (RR) is a technique that is often used for the purpose of benchmarking, and entails evicting content at random without taking its popularity or request rate into account. Generalizations to the RR scheme are proposed in \cite{tan2013optimal}, where the files are evicted with a certain probability that depends on their perceived popularity, tracked separately. We refer to the general scheme as the  the PRR scheme. For instance, a file that is not very popular will likely be evicted soon after it is added to the cache. Finally, other generalizations of the LRU-like algorithms include the meta-LRU and adaptive LRU algorithms discussed in \cite{li2017accurate}. Both algorithms maintain a virtual cache that records cache misses but does not necessarily store all files downloaded from the central server. 

While the eviction-based schemes were originally proposed for topology 2, they can also be adapted to topology 1. The key difference here is that the cache is not updated at every request but only periodically. The idea is to maintain a time-stamped list of cache misses. Whenever the cache update event is initiated, these new files replace the stored files as per the corresponding policy, e.g., with the least recently used files. 

\subsection{Proposed Storage Allocation policies} \label{proposed_techniques}

\begin{figure*}
	\centering
	\begin{tikzpicture}
	\draw[thick] (0,0)|-(1,0.5)|-(0,0) +(0.5,0)node[below]{10} +(0.5,0.5)node[above]{A} +(-0.25,-0.05)node[below]{$x_f:$} +(-0.25,0.45)node[above]{$f:$}
	(4,0.25) node[]{+} circle (0.25);
	\draw [thick,fill=red] (1,0)|-(1.5,0.5)|-(1,0) +(0.25,0)node[below]{8}  +(0.25,0.5)node[above]{B};
	\draw [thick,fill=green] (1.5,0)|-(2,0.5)|-(1.5,0) +(0.25,0)node[below]{6} +(0.25,0.5)node[above]{C};
	\draw [thick,fill=black] (2,0)|-(2.5,0.5)|-(2,0) +(0.25,0)node[below]{3} +(0.25,0.5)node[above]{D};
	\draw [->,very thick] (2.75,0.25) -- (3.5,0.25); 
	\draw [->,very thick] (4,-1) -- (4,-0.25);
	\draw [thick,fill=blue] (3.75,-1.75)|- +(0.5,0.5) |- +(0,0) +(0.5,0.75)node[right]{$x_h$} +(0.5,0.25)node[right]{7} +(0,0.25)node[left]{E} +(0,0.75)node[left]{$h$};
	\draw [thick,fill=yellow] (3.75,-2.5)|- +(0.25,0.5) |- +(0,0)+(0.5,0.25)node[right]{5} +(0,0.25)node[left]{F};
	\draw [thick,fill=cyan] (3.75,-3.25)|- +(0.5,0.5) |- +(0,0)+(0.5,0.25)node[right]{4} +(0,0.25)node[left]{G};
	\draw [thick,fill=magenta] (3.75,-4)|- +(0.25,0.5) |- +(0,0)+(0.5,0.25)node[right]{2} +(0,0.25)node[left]{H};
	
	\draw [thick] (1.25,1.5) node[]{Current Cache};
	
	\draw [very thick,->] (4.5,0.25) -- +(2.5,1.5);
	\draw [thick] (5.75,0.9) node[rotate=30,above]{Top-$X$};
	\draw [very thick,->] (4.5,0.25) -- +(2.5,0);
	\draw [thick] (5.8,0.45) node[]{Least-$X$};
	\draw [very thick,->] (4.5,0.25) -- +(2.5,-1.5);
	\draw [thick] (5.75,-0.6) node[rotate=-30,above]{Least-$X_{\text{th}}$};
	
	\draw [thick] (7.25,1.5)|- +(1,0.5) |- +(0,0) +(0.5,0.5)node[above]{A};
	\draw [thick,fill=red] (8.25,1.5)|- +(0.5,0.5) |- +(0,0) +(0.25,0.5)node[above]{B};
	\draw [thick,fill=blue] (8.75,1.5)|- +(0.5,0.5) |- +(0,0) +(0.25,0.5)node[above]{E};
	\draw [thick,fill=green] (9.25,1.5)|- +(0.5,0.5) |- +(0,0) +(0.25,0.5)node[above]{C};
	
	\draw [thick] (7.25,0)|- +(1,0.5) |- +(0,0) +(0.5,0.5)node[above]{A};
	\draw [thick,fill=blue] (8.25,0)|- +(0.5,0.5) |- +(0,0) +(0.25,0.5)node[above]{E};
	\draw [thick,fill=yellow] (8.75,0)|- +(0.25,0.5) |- +(0,0) +(0.125,0.5)node[above]{F};
	\draw [thick,fill=cyan] (9,0)|- +(0.5,0.5) |- +(0,0) +(0.25,0.5)node[above]{G};
	\draw [thick,fill=magenta] (9.5,0)|- +(0.25,0.5) |- +(0,0) +(0.125,0.5)node[above]{H};
	
	\draw [thick] (7.25,-1.5)|- +(1,0.5) |- +(0,0) +(0.5,0.5)node[above]{A};
	\draw [thick,fill=red] (8.25,-1.5)|- +(0.5,0.5) |- +(0,0) +(0.25,0.5)node[above]{B};
	\draw [thick,fill=blue] (8.75,-1.5)|- +(0.5,0.5) |- +(0,0) +(0.25,0.5)node[above]{E};
	\draw [thick,fill=yellow] (9.25,-1.5)|- +(0.25,0.5) |- +(0,0) +(0.125,0.5)node[above]{F};
	
	\draw [very thick,dashed] (9.75,2.5) +(1.25,0) -- +(1.25,-4.5);
	\draw [thick] (7.25,-2.25) +(1.25,0) node[]{Updated Cache} (12.25,-2.25) +(1,0) node[]{Evicted Files}; 
	
	\draw[thick,fill=black] (12.25,1.5)|- +(0.5,0.5) |- +(0,0) +(0.25,0.5)node[above]{D};
	
	\draw[thick,fill=red] (12.25,0)|- +(0.5,0.5) |- +(0,0) +(0.25,0.5)node[above]{B};
	\draw[thick,fill=green] (13,0)|- +(0.5,0.5) |- +(0,0) +(0.25,0.5)node[above]{C};
	\draw[thick,fill=black] (13.75,0)|- +(0.5,0.5) |- +(0,0) +(0.25,0.5)node[above]{D};
	
	\draw[thick,fill=green] (12.25,-1.5)|- +(0.5,0.5) |- +(0,0) +(0.25,0.5)node[above]{C};
	\draw[thick,fill=black] (13,-1.5)|- +(0.5,0.5) |- +(0,0) +(0.25,0.5)node[above]{D};
	
	\draw[->] (0,-3.5) -- (0,-1);
	\draw[->] (0,-3.5) -- +(1.5,0) node[right]{Size(MB)};
	\draw[thick] (0,-1.75)|- +(1,0.5) |- +(0,0) (0,-2.5)|- +(0.5,0.5) |- +(0,0) (0,-3.25)|- +(0.25,0.5)|- +(0,0);
	\draw[dotted] (0.25,-3.25) -- +(0,-0.25) node[below]{1} (0.5,-2.5) -- +(0,-1) node[below]{2} (1,-1.75) -- +(0,-1.75) node[below]{4};	 
	\end{tikzpicture}
	\caption{Illustration of proposed cache updating policies with an example} \label{caching_ex}
\end{figure*}
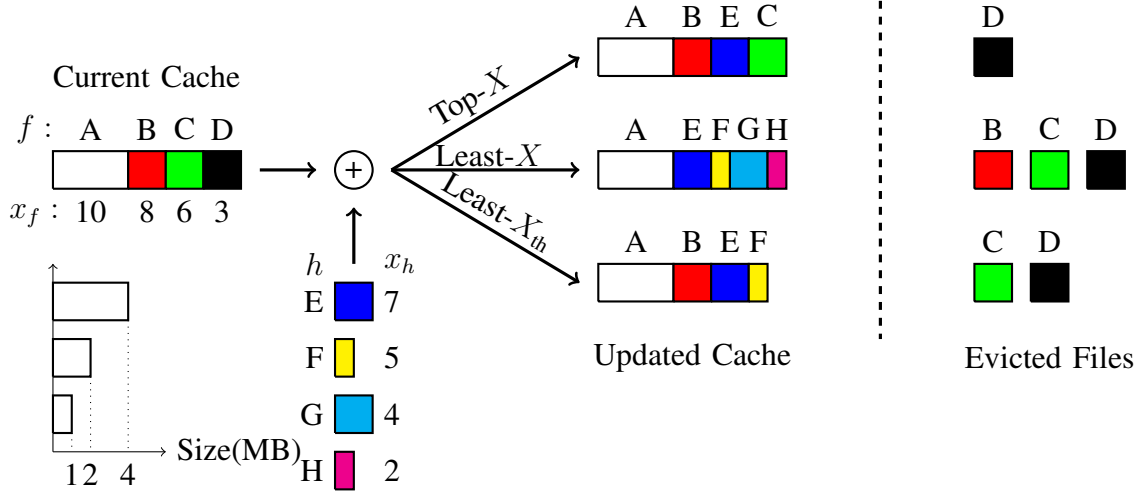
Different from the eviction-based schemes, the proposed techniques make use of the anticipated flows $\{x_{if}(t)\}$ to yield storage allocations $\m_i(t)$. Since the anticipated flows are designed to satisfy the \emph{average} instead of instantaneous demands, the resulting storage allocations would generally not include every file corresponding to a cache miss, depending on the value of $x_{if}(t)$. Such a policy is in fact a major departure from existing eviction-based algorithms such as LRU, where a file that misses the cache is always retained while already stored file is evicted. The implementation techniques applicable to the two topologies are discussed separately. 

Within the more general first topology, the content placement must occur periodically since the associated content fetching costs are high. The update interval, denoted by $T_{\text{up}}^v$, ranges from few hours to a day, and the files in the cache are updated at the start of each such interval. To this end, we propose three different content placement algorithms. 
	
\subsubsection{Top-$X$} Within the Top-$X$ scheme, the files with higher values of $x_{if}(t)$ are assumed to be popular and stored in the cache. That is, at each update event, the cache ensures that the files stored correspond to the ones with the largest values of $x_{if}(t)$. Since the file sizes may be different, the number of files stored may vary.  
	
	\subsubsection{Least-$X$ Used} 
	This technique seeks to combine LRU and the Top-$X$ algorithm described earlier. Specifically, at each update event, the files with the smallest values of $x_{if}(t)$ are replaced with the files that have recently been requested but were not available at the cache. If list of recently requested files is too large, only the files corresponding to the largest values of $x_{if}(t)$ are accommodated by respecting the cache storage constraints. 
	
	\subsubsection{Least-$X_{\text{th}}$ Used}
	Finally, the Least-$X_{\text{th}}$ algorithm maintains two different thresholds, for evicting and caching, respectively. At time $t$, let $\mathcal{F}_i(t)$ be the set of files cached at cache $i$, and $\mathcal{H}_i(t)$ be the set of files that were recently requested but were not available at the cache. Define thresholds $th_1 = \min_{f\in\mathcal{F}_i(t)}x_{if}(t)$  and $th_2 = \max_{h\in\mathcal{H}_i(t)}x_{ih}(t)$. A file $f \in \mathcal{F}_i(t)$ is evicted only when $x_{if}(t) < th_2$ while a file $h\in \mathcal{H}_i(t)$ is eligible for caching only if $x_{ih}(t) \geq th_1$. If the number of files in the set $\{h\in \mathcal{H}_i(t) | x_{ih}(t) \geq th_2\}$ is too large to be accommodated, only the files with the largest values of $x_{ih}(t)$ are cached.
	
	Fig. \ref{caching_ex} shows a toy example that serves to highlight the differences between the three proposed algorithms. At a given time, the cache server consists of files A, B, C, and D of sizes 4MB, 2 MB, 2 MB, and 2 MB, respectively. Within the previous slot, requests for files E, F, G, and H were received, none of which were available at the cache server. The $x$-values of the files at the start of the update process are shown in Fig. \ref{caching_ex}. As can be seen from the figure, the Top-$X$ algorithm is straightforward, and simply ensures that the files in the cache have the largest x values. The Least-$X$ algorithm is however different and takes into account the fact that the files E, F, G, and H have been recently requested. Consequently, all the recently requested files are accommodated while even those files with larger $x$ values but not recently requested (such as B and C)  are evicted. Finally, the Least-$X_{\text{th}}$ algorithm uses the thresholds $th_1 = 3$ and $th_2 = 7$. Files C and D are evicted as their x-values are below $th_2$. Likewise, files E, F, and G are eligible for storage. However, since not all of these can be accommodated, only the files with the largest x-values (namely, E and F) are cached.
	
	In the second topology, the files downloaded from the root server are available at the cache server without any extra cost. Consequently, the content placement and storage updates may occur per-file. While the algorithms, except the Top-$X$, may still be used for this case, we outline a fourth algorithm, termed as Least-$X_f$. Specifically, the Least-$X_f$ algorithm allows the cache $i$ to retain a file received from the root server only if $x_{if}(t) \geq \min_{f\in\mathcal{F}_i} x_{if}(t)$. In case the cache is full, the file with the smallest value of $x_{if}(t)$ is evicted. In other words, a file downloaded from the root server is cached as long as its popularity metric $x_{if}(t)$ is above that of any of the stored files. Such a policy is again very different from LRU, where a file downloaded from the root server is always stored. 

	\section{Simulation Results and Comparisons} \label{simulations}
This section provides numerical tests on synthetic and real data. The different file caching techniques are first evaluated and compared with state-of-the-art algorithms on real IPTV data. We begin with evaluating the performance of the content placement schemes detailed in Sec. III and IV. For simplicity, the setting consists of a single root and single cache server, since the proposed storage allocation algorithms are applied to each cache server separately. Interaction between cache servers, e.g., due to limited capacity of the root server, is not considered, and can be pursued later as co-operative caching. We emphasize the fact that all algorithms in the simulations are fully efficient in reaching user's demand, and hence the user's demands are always met. Root server has the back-haul links with sufficient capacity to serve the users. The algorithms here are mainly tested for their performance in tracking the dynamic popularity of a file in on-line fashion. 

The following performance criteria are utilized for the comparisons. 

\begin{figure}
	\setcounter{subfigure}{0}
	\begin{subfigure}{0.5\columnwidth}
		\includegraphics[width=\linewidth, height = 0.65\linewidth]
		{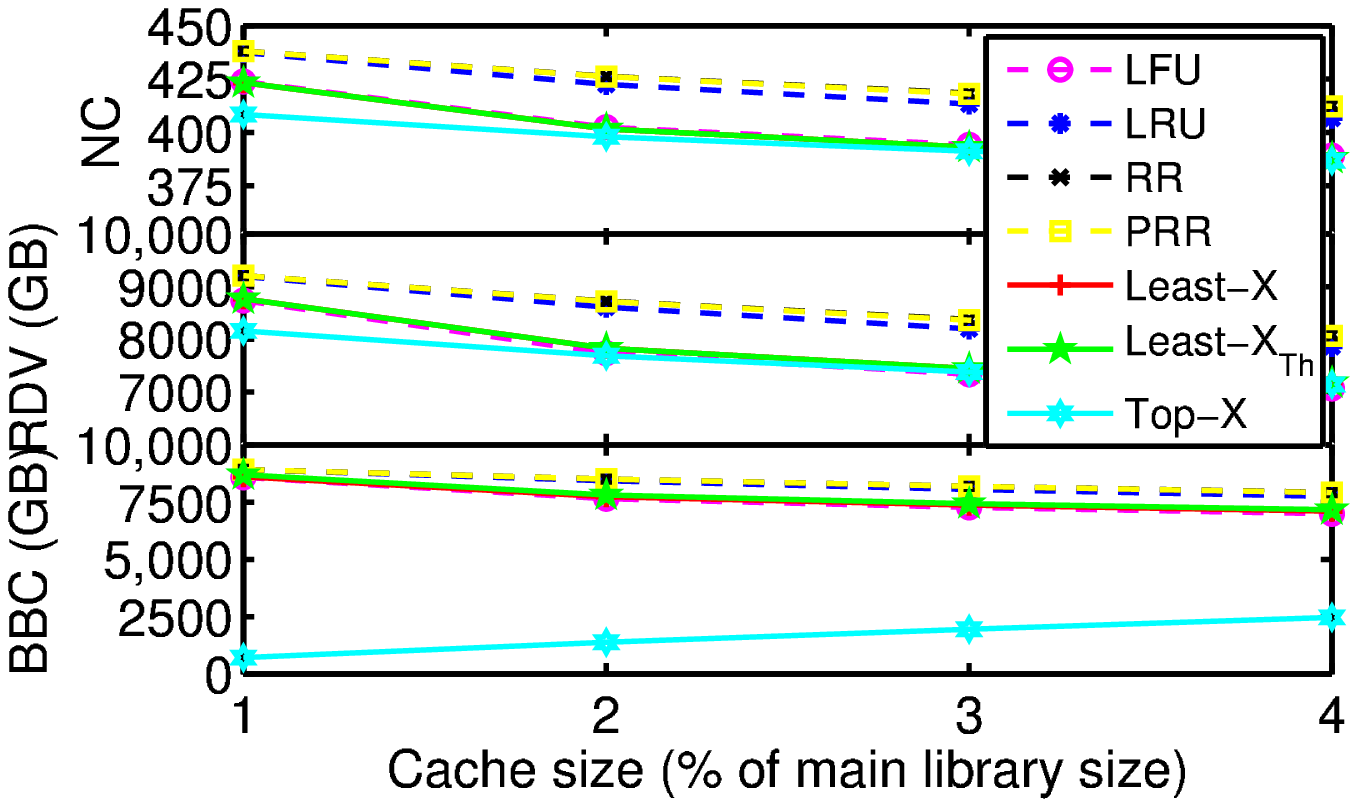}
		\caption{Topology 1}
		\label{Top1}
	\end{subfigure}
	\begin{subfigure}{0.5\columnwidth}
		\includegraphics[width=\linewidth,height = 0.65\linewidth]
		{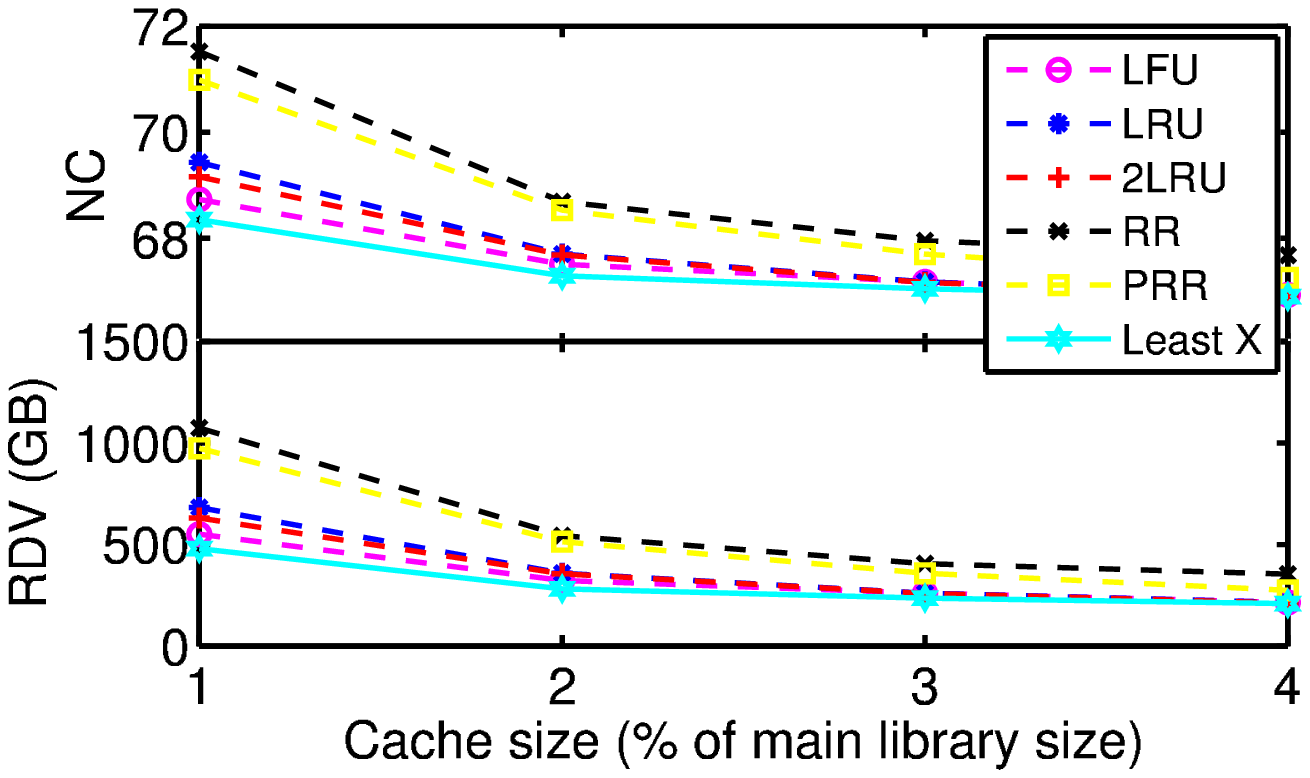}
		\caption{Topology 2}
		\label{Top2}
	\end{subfigure}
	\caption{Numerical results for synthetic data.}
	\label{synData}
	\vspace{-1.5cm}
\end{figure}

	\begin{itemize}
		\item \emph{Network Cost (NC)} is defined as in \eqref{cdnlin1} with cost functions $\chi_{i}(x)=(a_{ix}/2)x^2$ and $\varphi_{i}(y)=(a_{iy}/2)y^2$. As in realistic networks, the cost of serving content from the local server is lower than that of fetching the file from the root server, and we set $a_{ix}=1$, $a_{iy}=10$. A good caching scheme should result in a low average network cost.
		\item \emph{Rerouted Demand Volume (RDV)} is defined as an average amount of data that is requested from the cache but ultimately served by the root server. A good caching scheme should ensure that all demands are served by the cache and the rerouted demand volume is low. 
		\item \emph{Backhaul Bandwidth Consumption (BBC)} is defined as an average amount of data transferred between the cache server and the root server. Although the backhaul link between the cache and the root server is generally high-capacity, it is still preferred that the cache be stable and the placement algorithm should not request major placement changes every update interval. If popular files are cached in a timely manner, the backhaul consumption would be low.  
	\end{itemize}
Observe that the rerouted demand volume and the backhaul bandwidth consumption would be the same for Topology 2. Tests will be performed on the four state-of-the-art techniques (LRU, LFU, RR, and PRR) discussed in Sec. \ref{Existing_cache_techniques} and the three proposed techniques (Least-$X$, Top-$X$, Least-$X_{\text{th}}$, and Least-$X_f$) detailed in Sec. \ref{proposed_techniques}. 

\subsection{Static popularity} \label{staticsim}
We begin by first testing the performance on a synthetic demand data consisting of 400000 video files. The file sizes are chosen uniformly at random, ranging between 0.5 to 5 GB. The demands are determined by the file popularities that follow the Zipf distribution and remain constant over time. Specifically, the probability $p_k$ that the $k$-th most popular file is requested at a give time adheres to $p_k \propto k^{-s}$ where $s$ is the parameter that characterizes the skewness in the distribution. Such a model is suitable for archival videos and files whose popularity has reached a steady state. Zipf-like distribution is observed in real data traffic, and hence is widely used for testing content placement algorithms in the literature \cite{cha2009analyzing}. At each time slot, a user may demand a randomly chosen file with size uniformly distributed between 0.5 and 5 GB. The cache server receives demands from the users at an average rate of 4000 videos (1\% of content library) per hour and the experiment is conducted for 100 hours so as to allow the different algorithms to reach steady state. Note that the time slot duration here is chosen as 1 hour.
	
To begin with, we consider a worst-case scenario, where the cache server contains random files instead of the popular ones. As demands are received, the cache server runs the content placement algorithms and updates the files accordingly. For the purposes of the experiment, we utilize the first 10 hours of data for tuning the parameters $\mu$ and $T_{\text{up}}^{\lambda}$. Further, we set $T_{\text{up}}^v = T_{\text{up}}^{\lambda}$ since for the static case, consolidated demands received over an interval can be directly used to carry out the primal-dual updates and perform storage allocation. In other words, carrying out multiple primal-dual updates \eqref{ida}  per cache update interval does not help. Therefore, $T_{\text{up}}^{\lambda}$ does not affect the performance as long as it is less than the placement update interval. 

Fig. \ref{Top1} shows the performance of the proposed and state-of-the-art algorithms for Topology 1, as a function of the cache size. Since the simulation uses random file sizes, the cache size is expressed as percentage of the total size of main library. The network cost and rerouted demand volume are calculated for each hour (per time slot). As expected, both metrics improve as cache size is increased. Indeed, the performance of all the proposed algorithms is almost identical for larger cache sizes and matches with that of LFU. The behavior at lower cache sizes is more interesting, where it can be observed that the proposed algorithms outperform the state-of-the-art algorithms. Of these, the performance of RR is the worst, as expected, since it is agnostic to the file popularities. On the other hand, the PRR scheme where the file popularities are tracked, performs better. As also observed in \cite{garetto2016unified}, LFU performs the best under static popularity. Intuitively, in the static setting, the popular files continue to be requested for the entire duration of the simulation. Therefore it is better to keep the popular files in the cache and evict only the least frequently used rather than the least recently used ones. Interestingly, since the performance of the proposed algorithms is also based on tracking popular files, they outperform both LRU and LFU. The Top-$X$ scheme performs well in particular since like LFU, it tracks file popularities while discarding any information about how recently a particular file has been requested.

Within Topology 2, we also include the performance of the 2LRU algorithm, an LRU variant that maintains a virtual cache and is known to outperform LRU in some settings \cite{garetto2016unified}. Again the results for Topology 2 exhibit similar patterns as those in Topology 1, with LFU and Top-$X$ performing the best. In summary, for static settings, future demands can be better predicted by keeping track of the number of file requests rather than the time of these requests, as evident from the superior performance of Top-$X$ and LFU. It is remarked that for the sake of comparison, the cache sizes are deliberately kept small since the performance of all algorithms is almost the same for higher cache sizes. Indeed, the advantages of optimization-based algorithms become apparent only at lower cache sizes, as has also been observed earlier \cite{applegate2016optimal}. 

\subsection{Time-varying popularity}\label{simulations_IPTV}

	\begin{figure}
		\includegraphics[width=0.5\linewidth, height = 0.3\linewidth]
		{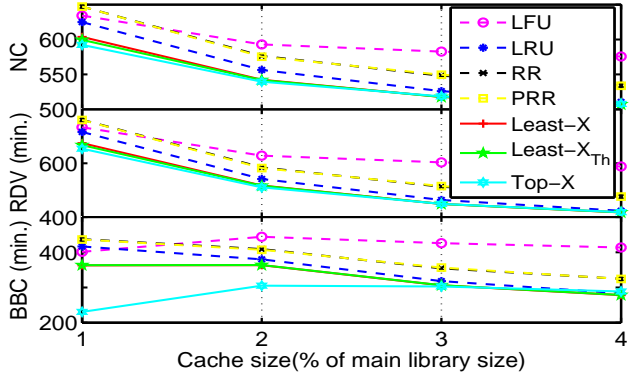}
		\caption{Performance metrics with $T_{\text{up}}^v$ = 6 hours.}
		\label{Real_6hr_Top_1}
	\end{figure}

	\begin{figure}
		\includegraphics[width=0.5\linewidth, height = 0.3\linewidth]
		{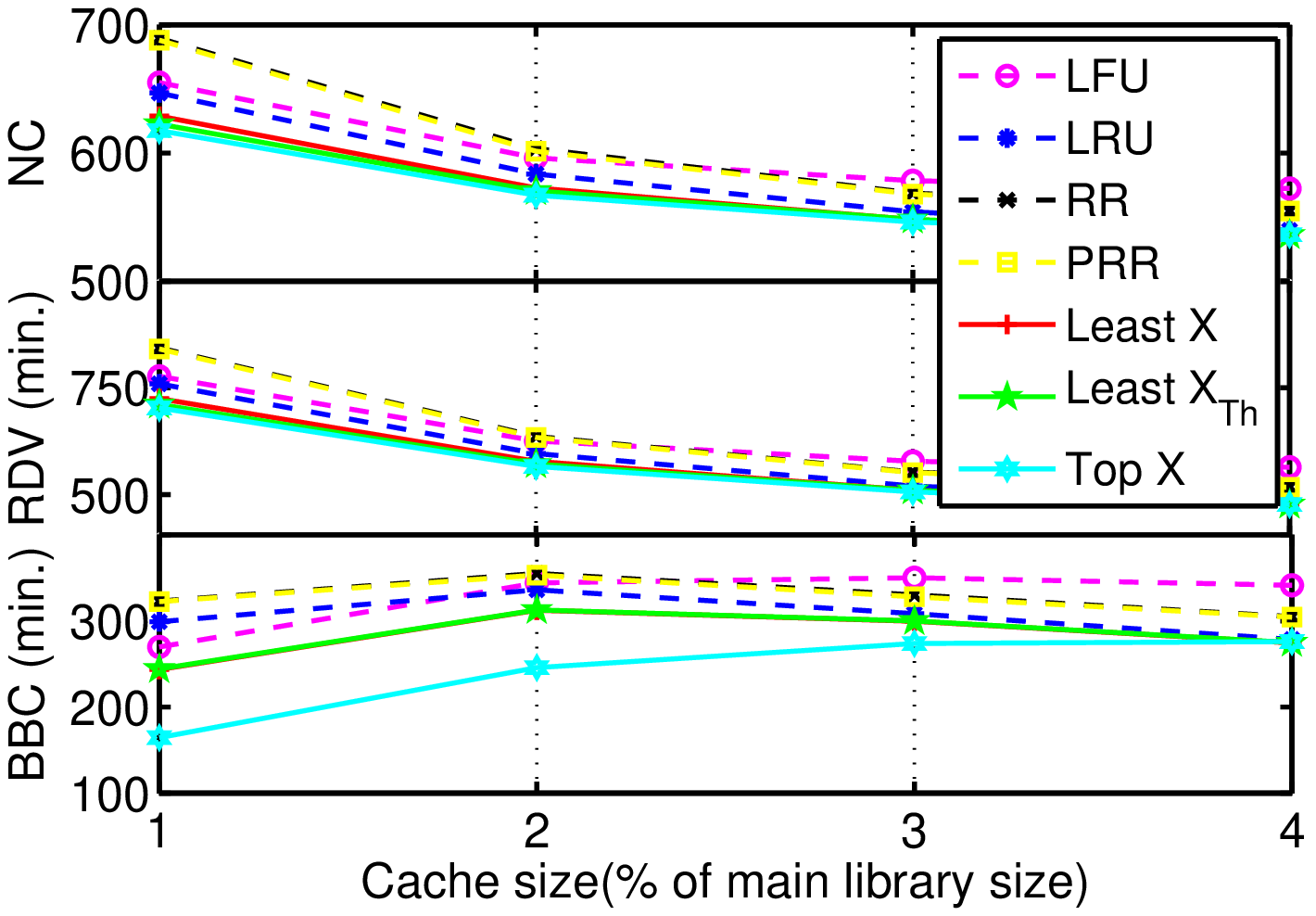}
		\caption{Performance metrics with $T_{\text{up}}^v$ = 12 hours.}
		\label{Real_12hr_Top_1}
	\end{figure}

	\begin{figure}
		\includegraphics[width=0.5\linewidth, height = 0.3\linewidth]
		{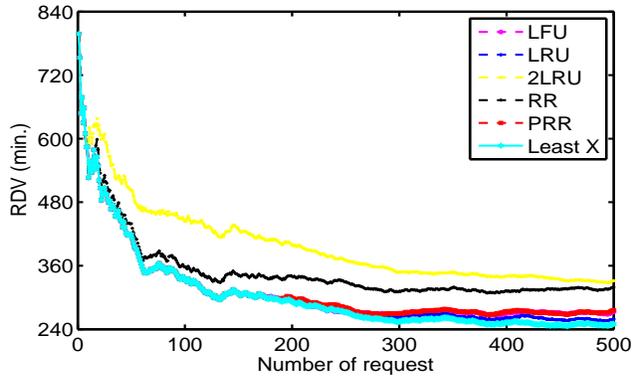}
		\caption{Running average of RDV over file requests with cache size = 1\%.}
		\label{Real_learn_rate}
	\end{figure}

	\begin{figure}
		\vspace{-0.5cm}
		\includegraphics[width=0.5\linewidth,height = 0.3\linewidth]
		{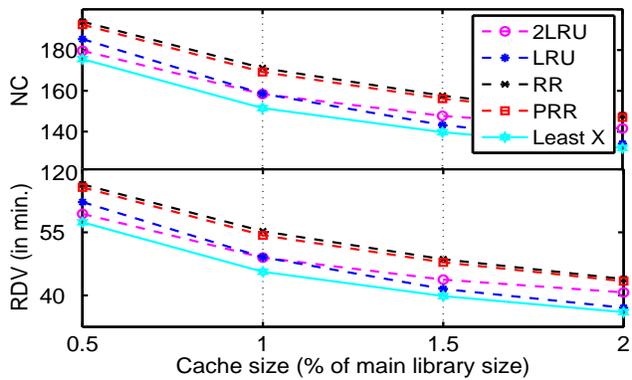}
		\caption{Performance for Topology 2}
		\label{Real_Top_2}
	\end{figure}

The IPTV dataset collected in  \cite{elkhatib2014dataset} is utilized to study the performance of the algorithm for realistic and time-varying demands. The dataset was collected in Lancaster Living Lab, which serves as a small IPTV service provider, and provides VoD and other services to its customers. The data contains the observations collected over a period of $7$ months from October 2011 to April 2012, and comprises of 44000 anonymized user demands from a central repository of 11000 unique files. Note however that no file size information is available in the dataset. To this end, we assume that the videos are encoded at a constant bit rate, and consequently, file sizes are a linear function of the video durations. Therefore, the performance metrics such as rerouted demand volume and bandwidth consumption are subsequently measured and reported in minutes instead of bits. We remark that, the IPTV service provider considered here may have had the video catalog for users. Hence one can predict the popularity accurately, which makes the content placement problem more trivial. The algorithms here are however tested under the assumption that no catalog is present to the users, and no algorithm has any idea of variation of popularity of any file.

Different from the synthetic dataset, the video files in the IPTV dataset exhibit dynamic popularities; specifically, the popularity of each file decays over time, while new files are regularly added. For the first topology, simulations are carried out for different cache update intervals, namely $T^v_{\text{up}} =6$, and $12$ hours. The parameters $\mu$ and $T_{\text{up}}^{\lambda}$ are tuned by considering the initial $2$ months of data and the experiments are conducted for the remaining $5$ months of data. The performance metrics for various proposed and state-of-the-art algorithms are shown in Figs. \ref{Real_6hr_Top_1},\ref{Real_12hr_Top_1}. The perceived popularities for files in PRR here are taken as the timestamps of the requests, similar to LRU. As evident from the plots, among the eviction-based algorithms, LRU exhibits the best while LFU exhibits the worst performance. Such a behavior is expected as LFU is agnostic to the time-varying nature of the popularity. Interestingly the proposed methods still outperform the eviction-based schemes especially for low cache sizes, resulting in low cost, rerouted demand volume, and backhaul bandwidth consumption. Recall from the updates that when requests for a particular file decrease, the value of $\lambda_{if}$ for that file also decays translating to lower anticipated flows $x_{if}$ and ultimately no storage allocation. In other words, the proposed algorithms, especially Top-$X$, are robust to temporal variations in the number of requests for each file, and continue to perform well regardless of the changes in content popularities. 

Recall that in the first topology, content placement is non-trivial and a decision to update the storage must be taken periodically. To this end, it is instructive to compare the performance of different algorithms for different update intervals $T_{\text{up}}^v$. As evident from Figs. \ref{Real_6hr_Top_1}, \ref{Real_12hr_Top_1}, both NC and RDV increase with $T_{\text{up}}^v$ while BBC decreases. That is, the parameter $T_{\text{up}}^v$ allows the service provider to trade-off running costs, arising from more frequent cache misses, against the periodic maintenance costs, arising from updating the cache. Interestingly however, the proposed algorithms continue to perform well for all parameter settings. Additionally, the performance of the proposed algorithms can be further tuned by changing the cost function, a feature not available to eviction-based algorithms. 

	\begin{figure}
		\includegraphics[width=0.5\linewidth, height = 0.3\linewidth]
		{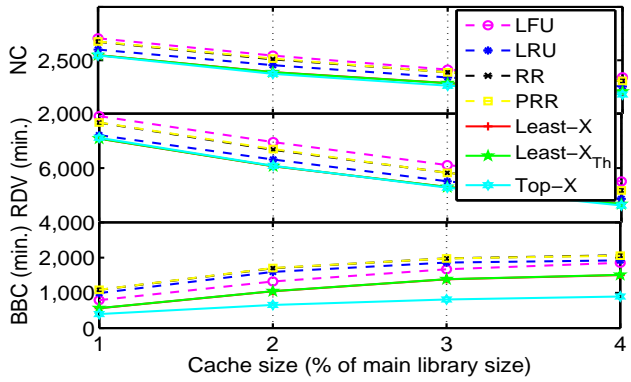}
		\caption{Performance metrics with $T_{\text{up}}^v$ = 6 hours.}
		\label{REAL_syn_6hr}
	\end{figure}

	\begin{figure}
		\includegraphics[width=0.5\linewidth,height = 0.3\linewidth]
		{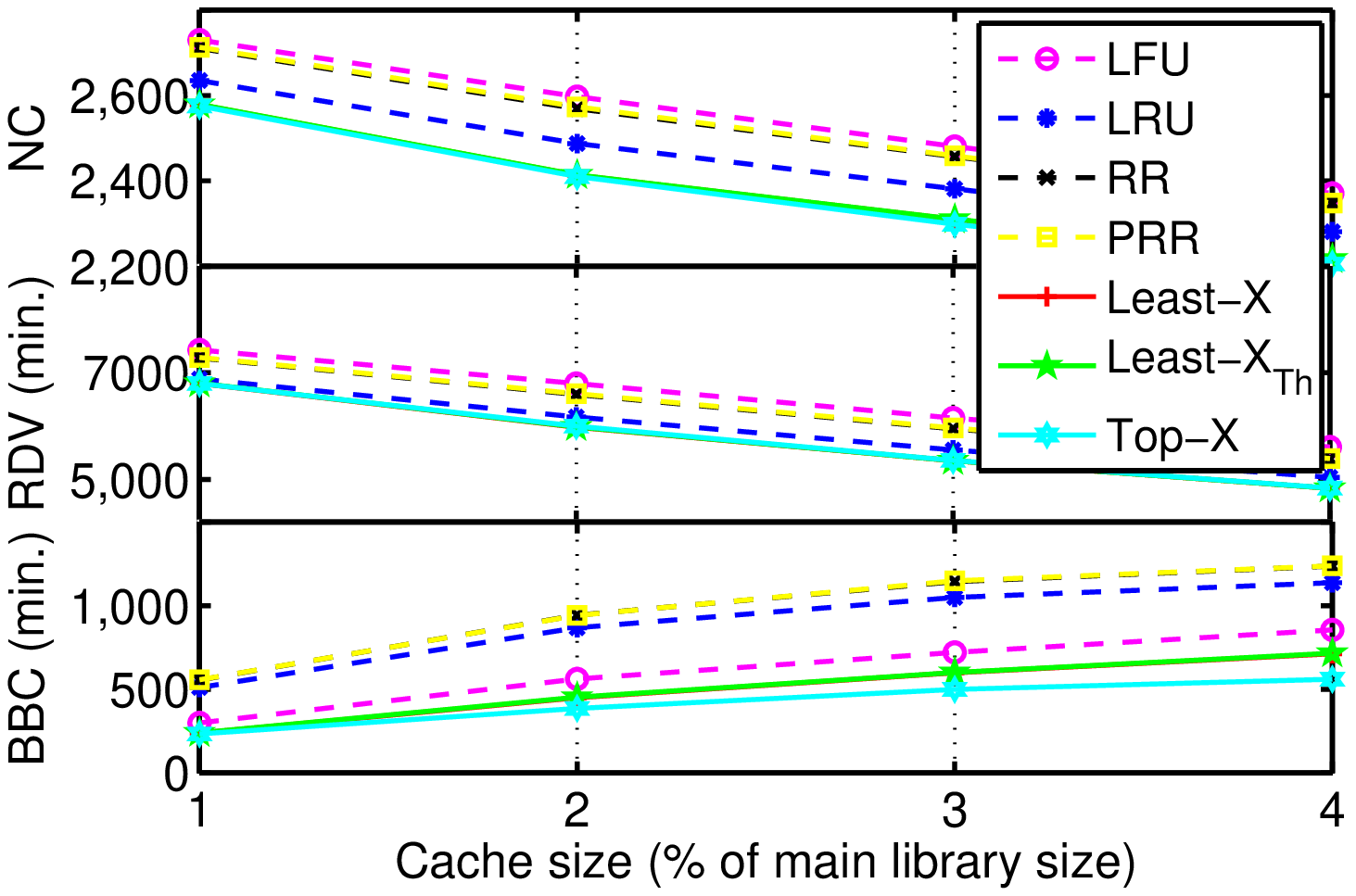}
		\caption{Performance metrics with $T_{\text{up}}^v$ = 12 hours.}
		\label{REAL_syn_12hr}
	\end{figure}

	\begin{figure}
		\vspace{-0.7cm}
		\includegraphics[width=0.5\linewidth,height = 0.3\linewidth]
		{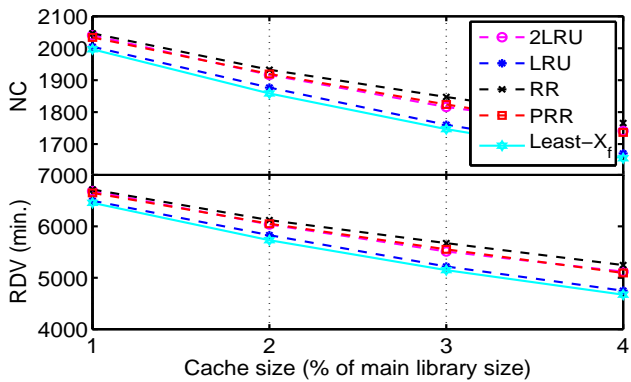}
		\caption{Performance for Topology 2}
		\label{REAL_syn_Top2}
	\end{figure}

Next, consider the second topology, where the performance of the proposed Least-$X_f$ algorithm is compared with that of the eviction-based algorithms. Fig. \ref{Real_Top_2} shows the performance metrics for the various cases. The proposed algorithm yields about 4\% reduction in the network cost and the rerouted demand volumes over LRU and a larger reduction over other algorithms. While the improvement is not significant, it demonstrates the universality of the proposed framework and the resulting low-complexity algorithms that not only outperform state-of-the-art techniques but are flexible enough to be applied to any topology with static or dynamic demands.

Before concluding, we also study the temporal evolution of the RDV for different algorithms, all starting from empty caches. As evident from Fig. \ref{Real_learn_rate}, the RDV of Least-$X_f$ is the lowest for all times. For the dataset under consideration, 2-LRU yields a higher steady-state RDV though it reaches its steady-state faster. It is remarked that since the popularities are time-varying, not much inference can be made regarding the 'steady-state' performance which continues to change over time. We also remark that, all algorithms are tested with out considering any delay to learn the popularity. Our main objective is to test the algorithms when the popularity varies dynamically in which learning delay would deteriorate the performance.

\subsection{Time-varying popularity for large-scale data}
The IPTV traces considered earlier are relatively small-scale when compared to modern large-scale content distribution services. In order to demonstrate the scalability of the proposed algorithms to very large-scale data, we utilize the technique in \cite{traverso2013temporal} to create a realistic statistical model of the data, and use it to generate an artificial dataset that is 10 times the size of the original dataset but follows a similar request distribution over files and time.

Specifically, each file $f$ is associated with the three parameters $(\tau_f, V_f, \ell_f(t))$ where $\tau_f$ is the time instant the first request for the file is made, $V_f$ is the total number of requests, and $\ell_f(t)$ is the time-dependent popularity profile for that file. For each file in the IPTV dataset, the quantity $\tau_f$ and $V_f$ are already known. We assume that $\ell_f(t)$ decays exponentially and estimate its decay rate using maximum likelihood estimation. In particular,  we have that $\ell_f(t) = \omega_f\exp\left(-\omega_ft\right)$ for $t\geq 0$ and zero otherwise, where we use the maximum likelihood estimate of the rate of decay $\omega_f$. Finally, the larger dataset is created by setting $V_f$ to be ten times the original value and resampling accordingly. Figs. \ref{REAL_syn_6hr}, \ref{REAL_syn_12hr}, \ref{REAL_syn_Top2} shows the results over the generated dataset. As can be seen, the performance of all algorithms follows approximately the same order with Top-$X$ outperforming. Interestingly, in Fig. \ref{REAL_syn_6hr} note that the increment in BBC as cache size increases which is, because of higher utilization of back-haul links. With large scale data, it happens that users are presented with wide variety of choices among the files. Hence when we operate the algorithms with in the small scale of cache sizes, cache hits are limited compared to the misses. That being said, it is natural that number of misses increases with size of cache, which makes the cache to download large number files.

	\section{Conclusion and future scope} \label{conclusion}

This work considered the problem of content placement in large scale CDNs. An optimization problem is formulated where the objective is to minimize the service-provider specific network cost function subject to capacity constraints on all the links. Since the user demands are not known a priori, dual decomposition is utilized to develop an online algorithm for the same. The proposed algorithm does not require independent identically distributed demands, and is guaranteed to yield a near-optimal solution to the relaxed problem. A primal variables obtained form the algorithm are subsequently used to make storage allocation decisions, while drawing inspiration from the existing eviction-based algorithms such as the least recently used (LRU) technique. Detailed numerical tests on synthetic and real IPTV data are provided and establish the efficacy of the proposed algorithms over other state-of-the-art algorithms. 

Before concluding, comments are provided on possible future directions. A limiting assumption in the current framework is that requests are not allowed to be rerouted to other caches as the demands aggregation occurred at the level of cache servers. In contrast, if requests may be rerouted, the optimization problem in \eqref{cdnint} must be appropriately modified for more general graph topologies than those in Fig. \ref{CDN_mainfig}. More importantly, the storage allocation would also not be straightforward since appropriate provisioning may be required for rerouted demands. The next generation of networks is expected to be more secure, scalable, and flexible via the adoption of content centric networking (CCN) \cite{ahlgren2012survey}. Going beyond the classical host-based network design, CCN routers are expected to have caching capabilities, thereby allowing efficient and scalable content delivery. The content placement problem for such arbitrary and possibly time-varying topologies is quite challenging and will be pursued as a future direction. 

	\appendices
	\section{Proof of Lemma \ref{primsol}}\label{aprimsol}
	For the purpose of this proof, define the strongly convex function $\gamma_f(u) := h(u)-\lambda_fu$ that appears in the objective of \eqref{pg}. 
	
	\subsubsection{Proof of Lemma \ref{primsol}(a)} By the way of contradiction, assume that $\hat{u}_f > 0$ so that there exists $\theta \in (0,1)$ such that $\theta \hat{u}_f + (1-\theta)g_f(0) = 0$. Since $g_f(0)$ is the unconstrained minimizer of $\gamma_f(u)$, i.e., $g_f(0) = \arg\min_u \gamma_f(u)$, it holds that $\gamma_f(\hat{u}_f) > \gamma_f(g_f(0))$. Therefore we have that
	\begin{align*}
	\gamma_f(0) & < \theta \gamma_f(\hat{u}_f) + (1-\theta)\gamma_f(g_f(0)) < \gamma_f(\hat{u}_f).
	\end{align*}
	In other words, replacing $\hat{u}_f$ with 0 yields a feasible solution of \eqref{pg} that attains a strictly lower objective function value. This is absurd, i.e., our original hypothesis that $u_f > 0$ is incorrect. It must therefore holds that $\hat{u}_f = 0$ whenever $g_f(0) < 0$. 
	\subsubsection{Proof of Lemma \ref{primsol}(b)} Again by the way of contradiction, assume that $\sum_f \hat{u}_f < C$ for the case when $\sum_f [g_f(0)]_{+} > C$. Note that since 
	\begin{align}\label{gf0}
	\{[g_f(0)]_{+}\}_f &= \arg \min_{\u \geq 0} \sum_f \gamma_f(u_f),
	\end{align}
	it holds that $\sum_f \gamma_f(\hat{u}_f) \geq \sum_f \gamma_f ([g_f(0)]_{+})$ since $\{\hat{u}_f\}$ also adheres to the constraint $\mathbf{1}^T\hat{\u} \leq C$. Proceeding as in part (a), it can be seen that there exists $\theta \in (0,1)$ such that $\theta \sum_f \hat{u}_f + (1-\theta) \sum_f [g_f(0)]_{+}  = C$. Denoting $\tilde{u}_f := \theta \hat{u}_f + (1-\theta) [g_f(0)]_{+} \geq 0$ for all $f\in \mathcal{F}$, it follows from the convexity of $\gamma_f(\cdot)$ that
	\begin{align*}
	\sum_f\gamma_f(\tilde{u}_f) < \theta \sum_f\gamma_f(\hat{u}_f) + (1-\theta)\sum_f\gamma_f([g_f(0)]_{+}) < \sum_f\gamma_f(\hat{u}_f)
	\end{align*}
	That is, the point $\{\tilde{u}_f\}$ adheres to the constraints in \eqref{pg}, while achieving a lower objective value than the optimum $\{\hat{u}_f\}$. Such a conclusion is absurd, implying that the original hypothesis that $\sum_f \hat{u}_f < C$ is incorrect. It must therefore hold that $\sum_f \hat{u}_f = C$ whenever $\sum_f [g_f(0)]_{+} > C$.
	
	\subsubsection{Proof of Lemma \ref{primsol}(c)} Associate Lagrange multiplier $\upsilon$ with the constraint $\mathbf{1}^T\u \leq C$ and observe that the primal-dual solution $(\hat{\u},\hat{\upsilon})$ must satisfy the KKT conditions, namely, the optimality condition $\hat{\u} = \arg \min_{\u\geq 0} \sum_f \gamma_f(u_f) + \hat{\upsilon} u_f$, complementary slackness $\hat{\upsilon}(\mathbf{1}^T\u-C) = 0$, and the feasibility conditions $\mathbf{1}^T\hat{\u} \leq C$ and $\hat{\upsilon} \geq 0$. The proof follows by verifying these conditions for \eqref{pgsol}. In the case when $\sum_f [g_f(0)]_{+} \leq  C$, it is straightforward to see from \eqref{gf0} that the pair $(\{[g_f(0)]_{+}\}_f, 0)$ satisfies the KKT conditions. For the second case, observe from the definition of $\hat{\upsilon}$ in \eqref{pgsol} that the pair $(\{[g_f(\hat{\upsilon})]_{+}\}_f, \hat{\upsilon})$ satisfies the feasibility and the complementary slackness conditions. The final KKT condition also follows from the observation that 
	\begin{equation}\label{gmu}
	[g_f(\hat{\upsilon})]_{+} = \arg \min_{u \geq 0} h(u) - (\lambda_f - \hat{\upsilon})u  \nonumber 
	\end{equation}
	Since $h(\cdot)$ is strongly convex, the function $\sum_f g_f(z)$ is monotonically decreasing in $z$, so that the solution to $\sum_fg_f(\hat{\upsilon}) = C$ is unique. Finally, since $\mathbf{1}^T\hat{\u} = C$, there must exist some $f\in \mathcal{F}$ such that $\hat{u}_f > 0$. This implies that there exists $f$ such that $g_f(\hat{\upsilon}) > 0$ or equivalently $\hat{\upsilon} < \lambda_f - h'(0)$ for some $f\in \mathcal{F}$, which yields the required range for $\hat{\upsilon}$. 
	
	\section{Proof of Lemma \ref{lambda_bound} }\label{alamb}
	We begin by substituting $x_2 = y_2 = 0$, $x_1 = \hat{x}_{if}(t)$, and $y_1 = \hat{y}_{if}(t)$ in \eqref{a3}, to obtain
	\begin{subequations} \label{lc}
		\begin{align}
		m_\chi \hat{x}_{if}(t) \leq \chi_i'(\hat{x}_{if}(t))-\chi_i'(0) \leq L_\chi \hat{x}_{if}(t) \label{xCond}\\
		m_\varphi \hat{y}_{if}(t) \leq \varphi_i'(\hat{y}_{if}(t))-\varphi'_i(0) \leq L_\varphi \hat{y}_{if}(t). \label{yCond}
		\end{align}
	\end{subequations}
	The bounds in \eqref{boundlambdat} will be established by induction. For the base case of $t= 0$, the bounds already hold from \eqref{lambdalimits}. Suppose that the bounds in \eqref{boundlambdat} also apply for some $t\geq 0$. It remains to prove that the bounds continue to hold for $t+1$. 
	
	To this end, define $\phi := \min\{\chi_i'(0), \varphi_i'(0)\}$ and $\psi := \max\{\chi_i'(0), \varphi_i'(0)\}$. Also, recall from Lemma \ref{primsol} that since $g_f(\upsilon)$ is a monotonically decreasing function of $\upsilon$, it holds that
	\begin{align}\label{xy}
	0 \leq \hat{x}_{if}(t) \leq [\chi_i'^{-1}(\lambda_{if}(t))]_{+}, \hspace{1cm}0 \leq \hat{y}_{if}(t) \leq [\varphi_i'^{-1}(\lambda_{if}(t))]_{+}.
	\end{align}
	The proof is divided into two parts, corresponding to \eqref{xCond} and \eqref{yCond}. 
	
	\subsection{Proof of \eqref{minlambda}:}  By induction hypothesis, we have that $\lambda_{if}(t) \geq \phi$, and proof is divided into two separate cases:
	
	\noindent \textbf{Case 1. } $\lambda_{if}(t) \geq \psi$, which implies that
	\begin{align}\label{xy2}
	\chi_i'^{-1}(\lambda_{if}(t)) \geq 0, \hspace{1cm} \varphi_i'^{-1}(\lambda_{if}(t)) \geq 0
	\end{align}
	Using the fact that $\chi_i'$ and $\varphi_i'$ are monotonic, and substituting \eqref{xy2} into \eqref{xy}, we obtain
	\begin{align}\label{lamineq}  
	\chi_i'(\hat{x}_{if}(t)) \leq \lambda_{if}(t), \hspace{1cm} \varphi_i'(\hat{y}_{if}(t)) \leq \lambda_{if}(t) 
	\end{align}
	Next, using \eqref{lc} and \eqref{lamineq} we obtain the following series of inequalities:
	\begin{align*}
	\hat{x}_{if}(t) + \hat{y}_{if}(t) &\leq \frac{\chi_i'(\hat{x}_{if}(t)) - \chi_i'(0)}{m_\chi} + \frac{\varphi_i'(\hat{y}_{if}(t)) - \varphi_i'(0)}{m_\varphi}  \\
	&\leq \frac{\lambda_{if}(t) - \chi_i'(0)}{m_\chi} + \frac{\lambda_{if}(t) - \varphi_i'(0)}{m_\varphi} \\
	&\leq\frac{\lambda_{if}(t)-\phi}{m} 
	\end{align*}
	where the last inequality follows from the definitions of $\phi$ and $m$. Therefore, the use of the induction hypothesis yields
	\begin{align*}
	\lambda_{if}(t+1) &= \lambda_{if}(t) - \mu(\hat{x}_{if}(t)+\hat{y}_{if}(t)-d_{if}(t)) \\
	&\geq \left(1-\frac{\mu}{m}\right)\lambda_{if}(t) + \frac{\mu}{m}\phi + \mu d_{if}(t) \geq \phi 
	\end{align*}
	
	\noindent \textbf{Case 2. } $\lambda_{if}(t) < \psi$. For the sake of clarity, lets assume with out loss of generality, $\psi=\varphi_i'(0)$. From the statement of Lemma \ref{primsol}, one can deduce that, $\hat{y}_{if}(t)=0$. By following similar steps as in \textbf{Case 1} for $\hat{x}_{if}(t)$, we can write
	\begin{align*}
	\hat{x}_{if}(t) + \hat{y}_{if}(t) \leq\frac{\lambda_{if}(t)-\phi}{m_{\chi}}
	\end{align*}
	Finally we can conclude by saying,
	\begin{align*}
	\lambda_{if}(t+1) \geq \left(1-\frac{\mu}{m_{\chi}}\right)\lambda_{if}(t) + \frac{\mu}{m_{\chi}}\phi + \mu d_{if}(t) \geq \phi 
	\end{align*} 
	which is the desired result. 
	
	\vspace{-0.35cm}
	\subsection{Proof of \eqref{maxlambda}: } From the induction hypothesis, we have that $\sum_{f}\lambda_{if}(t) \leq \Delta$. This part of the proof is divided into three separate cases:
	
	\noindent \textbf{Case 1. } $\sum_f \hat{x}_{if}(t) < C^c_i$ and $\sum_f \hat{y}_{if}(t) < C^r_i$: In this case, it holds that $\hat{x}_{if}(t) = [\chi_i^{-1}(\lambda_{if}(t))]_{+} \geq \chi_i^{-1}(\lambda_{if}(t))$. Again using the fact that $\chi_i'$ is an increasing function, it follows from \eqref{lc} that
	\begin{align}
	\hat{x}_{if}(t) \geq \frac{\lambda_{if}(t)-\chi'(0)}{L_\chi} \geq \frac{\lambda_{if}(t)-\psi}{L_\chi}
	\end{align}
	Taking sum over all $f \in \mathcal{F}$, we obtain 
	\begin{align}
	\sum_f \hat{x}_{if}(t) &\geq \frac{\sum_f \lambda_{if}(t)-F\psi}{L_{\chi}} 
	\end{align}
	Likewise, since $\sum_f \hat{y}_{if}(t) < C^r_i$, it follows from \eqref{lc} that $\sum_f \hat{y}_{if}(t) \geq (\sum_f \lambda_{if}(t)-F\psi)/L_\varphi$. The induction hypothesis therefore yields:
	\begin{align*}
	\sum_{f} \lambda_{if}(t+1) &= \sum_f\lambda_{if}(t) - \mu\left(\sum_{f} \hat{x}_{if}(t) + \hat{y}_{if}(t)\right)+ \mu \sum_f d_{if}(t) \\
	&\leq\! \left(\!1-\frac{\mu}{L}\right)\! \sum_{f}\! \lambda_{if}(t)\! +\! \frac{\mu F \psi}{L}\!  +\! \mu \sum_f\! d_{if}(t) \\
	& \leq \!\left(\!1-\frac{\mu}{L}\right)\! \Delta + \frac{\mu (F\psi+(C^c_i + C^r_i)L) }{L} \leq \Delta
	\end{align*}
	
	\noindent \textbf{Case 2. } $\sum_f \hat{x}_{if}(t) = C^c_i$ and $\sum_f \hat{y}_{if}(t) = C^r_i$: in this case, it directly follows from the induction hypothesis that
	\begin{align*}
	\sum_{f} \lambda_{if}(t+1) &=\sum_f\lambda_{if}(t) - \mu\left(\sum_{f} \hat{x}_{if}(t) + \hat{y}_{if}(t)\right)+ \mu \sum_f d_{if}(t) \\
	&\! \leq \!\sum_f\!\lambda_{if}(t)\! - \mu C^c_i\! - \mu C^r_i\!  + \mu C_i^c\! +\mu C_i^r\! \leq  \Delta 
	\end{align*}
	
	\noindent \textbf{Case 3. } $\sum_f \hat{x}_{if}(t) < C^c_i$ and $\sum_f \hat{y}_{if}(t) = C^r_i$: similar to cases 1 and 2, it holds that
	\begin{align*}
	\sum_{f} \lambda_{if}(t+1) &=\sum_f\lambda_{if}(t) - \mu\left(\sum_{f} \hat{x}_{if}(t) + \hat{y}_{if}(t)\right)+ \mu \sum_f d_{if}(t) \\
	&\leq (1-\frac{\mu}{L_\chi}) \Delta + \frac{\mu (F\psi+C^c_iL_\chi) }{L_\chi} \leq \Delta
	\end{align*}
	
	\noindent \textbf{Case 4. } $\sum_f \hat{x}_{if}(t) = C^c_i$ and $\sum_f \hat{y}_{if}(t) < C^r_i$: similar to case 3, it holds that
	\begin{align*}
	\sum_{f} \lambda_{if}(t+1) &\leq (1-\frac{\mu}{L_\varphi}) \Delta + \frac{\mu (F\psi+C^r_iL_\varphi) }{L_\varphi} \leq \Delta 
	\end{align*}

	\section{Proof of Theorem \ref{thm}}\label{athm}
	First consider the subproblems in \eqref{xup} and \eqref{yup}. Associate Lagrange multipliers ${\alpha}_i(t)$ and ${\xi}_{if}(t)$ with the capacity and the non-negativity constraints, respectively, in \eqref{xup}. Likewise associate Lagrange multipliers ${\beta}_i(t)$ and ${\zeta}_{if}(t)$ with the constraints in \eqref{yup}. Denoting the optimal dual variables for \eqref{xup}-\eqref{yup} by $\hat{\alpha}_i(t)$, $\hat{\xi}_{if}(t)$, $\hat{\beta}_i(t)$, and $\hat{\zeta}_{if}(t)$, it follows from the KKT conditions that
	\begin{enumerate}
		\item[K1. ] Primal feasibility: $\sum_f \hat{x}_{if}(t) \leq C^c_i$, $\hat{x}_{if}(t) \geq 0$, $\sum_f \hat{y}_{if}(t) \leq C^r_i$, and  $\hat{y}_{if}(t) \geq 0$
		\item[K2. ] Dual feasibility: $\hat{\alpha}_i(t), \hat{\xi}_{if}(t), \hat{\beta}_i(t), \hat{\zeta}_{if}(t) \geq 0$
		\item[K3. ] Complementary slackness: $\hat{\alpha}_i(t)(\sum_f \hat{x}_{if}(t) - C^c_i) = \hat{\xi}_{if}(t)\hat{x}_{if}(t) = \hat{\beta}_i(t)(\sum_f \hat{y}_{if}(t) - C^r_i) = \hat{\zeta}_{if}(t)\hat{y}_{if}(t) = 0$
		\item[K4. ] First order optimality condition:
		\begin{align*}
		\hat{\alpha}_i(t) + \chi_i'(\hat{x}_{if(t)}) - \lambda_{if}(t) - \hat{\xi}_{if}(t) = 0, \hspace{0.5cm}\hat{\beta}_i(t) +  \hat{\rho}_i'(\hat{y}_{if(t)}) - \lambda_{if}(t) - \hat{\zeta}_{if}(t) =0  \hspace{0.25cm}\forall ~ i,f
		\end{align*}
	\end{enumerate} 
	
	Based on the conditions (K1)-(K4), it follows that $\hat{x}_{if}(t)$ and $\hat{y}_{if}(t)$ are primal feasible for the capacity and non-negativity constraints of \eqref{cdnlin} for all $t \geq 0$. Likewise, it follows from (K1)-(K4) that $\{\hat{\alpha}_i(t), \hat{\beta}_i(t), \hat{\xi}_{if}(t), \hat{\zeta}_{if}(t)\}_{t=1}^T$ also satisfy the dual feasibility and complementary slackness conditions for \eqref{cdnlin}. The rest of the proof is devoted to showing \eqref{thmpf}-\eqref{thmlt}. 
	
	Towards showing \eqref{thmpf}, we begin with taking summation over $t = 0$, $\ldots$, $T-1$ in \eqref{lup} and canceling the common terms on both sides, which yields
	\begin{align*}
	\lambda_{if}(T) &=  \lambda_{if}(0) -\mu \sum_{t=0}^{T-1} \left(x_{if}(t) + y_{if}(t)- d_{if}(t) \right) \nonumber\\
	\Rightarrow ~~~ \frac{\lambda_{if}(T)-\lambda_{if}(0)}{T} &= -\frac{\mu}{T} \sum_{t=0}^{T-1}\left(x_{if}(t) + y_{if}(t) - d_{if}(t) \right). \label{telesum}
	\end{align*}
	From Lemma \ref{lambda_bound}, we have that $\lambda_{if}(T)-\lambda_{if}(0) \in [-\Delta, \Delta]$. Therefore, it follows that 
	\begin{align}
	\left|\frac{1}{T}\sum_{t=0}^{T-1}\left(x_{if}(t) + y_{if}(t) - d_{if}(t) \right)\right| \leq \mathcal{O}\left(\frac{1}{\mu T}\right)
	\end{align}
	
	Next, for \eqref{thmlt}, we begin with the observation that 
	\begin{align*}
	\nabla_{x_{if}(t)}L_T(\{\hat{\Z}(t)\}_{t=1}^T, \nu_{if}) = \frac{1}{T}\left(\hat{\alpha}_i(t) + \chi_i'(\hat{x}_{if(t)}) - \nu_{if} - \hat{\xi}_{if}(t)\right) = \frac{\lambda_{if}(t)-\nu_{if}}{T}
	\end{align*}
	Likewise, it holds that $\nabla_{y_{if}(t)} L_T(\{\hat{\Z}(t)\}_{t=1}^T, \nu_{if}) = \left(\lambda_{if}(t)-\nu_{if}\right)/T$. Therefore from Lemma \ref{lambda_bound} it follows that
	\begin{align*}
	\left\|\nabla_{\{\X(t),\Y(t)\}_{t=1}^T} L_T(\{\hat{\Z}(t)\}_{t=1}^T, \nu_{if})\right\|^2 =  \frac{4}{T^2}\sum_{i=1}^N\sum_{f=1}^F \sum_{t=1}^T(\lambda_{if}(t)-\nu_{if})^2
	\end{align*}
	The value of $\{\nu_{if}\}_{i,f}$ that minimizes this bound is given by $\nu_{if} = \frac{1}{T}\sum_{t=1}^T \lambda_{if}(t)$. Since $\lambda_{if}(t) \leq \Delta$, it therefore holds that
	\begin{align*}
	\left\|\nabla_{\{\X(t),\Y(t)\}_{t=1}^T}  L_T(\{\hat{\Z}(t)\}_{t=1}^T, \frac{1}{T} \sum_{t=1}^T  \lambda_{if}(t))\right\| \leq \frac{2\sqrt{NF}\Delta}{\sqrt{T}}
	\end{align*}

	\footnotesize
	\bibliographystyle{IEEEtran}
	\bibliography{IEEEabrv,references}

\end{document}